\documentclass[useAMS,usenatbib,fleqn]{mnras}
\newif\ifAMStwofonts
\AMStwofontstrue

\usepackage{amsmath}	
\usepackage{amssymb}	
\usepackage{nccmath}

\def\pmb#1{\mbox{\boldmath$#1$}}
\def\gtsim {>\kern-1.2em\lower1.1ex\hbox{$\sim$}}
\def\ltsim {<\kern-1.2em\lower1.1ex\hbox{$\sim$}}
\def\gtsim {>\kern-1.2em\lower1.1ex\hbox{$\sim$}}
\def\ltsim {<\kern-1.2em\lower1.1ex\hbox{$\sim$}}

\def\be{\begin{equation}}
\def\ee{\end{equation}}
\def\be{\begin{eqnarray}}
\def\ee{\end{eqnarray}}

\usepackage{color}

\def\pmbmt#1{\pmb{\sf #1}}
\def\rmi{{\rm i}}



\usepackage{graphicx}
\usepackage{epsfig}
\begin{document}

\title{Non-linear excitation of low frequency modes by overstable convective modes
in rotating stars}
\author[U. Lee]{ Umin Lee$^1$\thanks{E-mail: lee@astr.tohoku.ac.jp}
\\$^1$Astronomical Institute, Tohoku University, Sendai, Miyagi 980-8578, Japan}

\date{Typeset \today ; Received / Accepted}
\maketitle


\begin{abstract}
We discuss non-linear excitation and amplitude saturation of $g$-modes, $r$-modes and
overstable convective (OsC) modes in early type main sequence stars,
taking account of the effects of three-mode couplings on amplitude evolutions.
OsC modes are rotationally stabilized convective modes in the convective core and they resonantly excite
low frequency $g$-modes to obtain large amplitudes in the envelope when the rotation rate of the core is larger than critical rates.
We use, for a network of three-mode couplings, amplitude equations governing the time evolution of the mode amplitudes where each of three-mode couplings is assumed to occur between two stable modes and one unstable mode.
Assuming that the unstable modes in the couplings are OsC modes in the core
and the stable modes are $g$- and $r$-modes in the envelope, 
we integrate the amplitude equations to see how the $g$- and $r$-modes are non-linearly excited by the OsC modes and 
whether or not the amplitude evolutions tend toward a state of finite amplitudes.
We find that the non-linear three-mode couplings do excite low frequency $g$- and $r$-modes but they are not necessarily effective to achieve amplitude saturation
since the three-mode couplings between the OsC modes with large growth rates and $g$- and $r$-modes with
small damping rates tend to destabilize amplitude evolutions.

\end{abstract}



\section{Introduction}

For self-excited stellar pulsations, non-linearities are essential 
for saturation of their amplitude growth, as shown in dynamical simulations of radial pulsators (e.g., \citealt{Stellingwerf1975}).
For non-radial pulsations of rotating stars, however, it is not an easy task to carry out
dynamical simulations of pulsating stars to estimate the amplitudes at saturation (e.g., \citealt{LindblomTohlineVallisneri2001,LindblomTohlineVallisneri2002} for the $r$-modes in rotating neutron stars).
A usual step we can take to estimate the amplitudes of non-radial oscillations at saturation may be to take into considerations 
the effects of non-linear mode-couplings between linear non-radial modes on their
amplitude growth and saturation.
Non-linear couplings between linear waves have been a topic extensively investigated in fluid mechanics
(e.g., \citealt{Craik1985}).

\citet{Dziembowski1982} and \citet{DziembowskiKrolikowska1985} may be among the earliest attempts in which
the lowest order non-linear mod-couplings are applied to small amplitude non-radial oscillations of stars 
to discuss the parametric excitation and amplitude saturation of the oscillations.
For non-linear three-mode couplings, \citet{Dziembowski1982} derived the amplitude equations which contain up to quadratic terms in amplitudes to describe amplitude evolutions, 
assuming that the linear adiabatic eigenfunctions form a complete set of functions.

Non-linear three-mode couplings between stellar oscillations have been applied to various targets
to explain the excitation and saturation of the oscillation modes, additional damping effects in stars, 
and so on.
For example, assuming that the tides in stars in a binary system are coupled non-linearly to oscillation modes of the stars, \citet{KumarGoodman1996} discussed additional damping effects of the tides on the binary evolution. 
\citet{Weinberg_etal2012} studied the excitation and damping of tides in close binary systems to 
discuss non-linear corrections to linear tidal theory.
\citet{Arras_etal2003}, on the other hand, used the effects of non-linear couplings with inertial modes 
to estimate the saturation amplitudes of the $r$-modes destabilized by gravitational wave emissions from rotating neutron stars (see also \citealt{BrinkTeukolskyWasserman2005,BondarescuWasserman2013}).
\citet{WuGoldreich2001} assumed parametric instability of $g$-modes in hydrogen white dwarfs to 
theoretically obtain mode amplitudes consistent with observations where
unstable $g$-modes are assumed to suffer convective driving (\citealt{Brickhill1990,Brickhill1991}).

There are pulsating variables for which the excitation mechanisms for pulsations are not necessarily
well understood.
For example, \citet{Saio_etal2018} have attributed low frequency light variations found in $\gamma$ Dor stars,
spotted stars and so called heart beat stars to $r$-mode oscillations, but
excitation mechanisms for the $r$-modes are not necessarily identified.
Note however that a recent paper by 
\citet{SaioKurtz2022} have suggested that the $r$-modes in many hear beat stars are tidally excited by the
companion stars.
It is also noticed that excitation mechanisms need to be identified 
for low frequency $g$-modes detected in early type stars in the main sequence stages between the $\gamma$ Dor stars and slowly pulsating B (SPB) stars (e.g., \citealt{Kahramanetal20} for hot $\gamma$ Dor stars) since neither the excitation mechanism called convective blocking for $\gamma$ Dor stars (e.g., \citealt{Guziketal2000,Dupretetal05}) nor the $\kappa$-mechanism for SPB stars (e.g., \citealt{GautschySaio1993,Dziembowskietal93}) is viable in the gap domain between the two classes of pulsating variables.

For an excitation mechanism of low frequency modes detected, for example, in hot $\gamma$ Dor stars, we examine in this paper non-linear excitation of low frequency $g$- and $r$-modes by overstable convective (OsC) modes in rotating early type main sequence stars.
We consider the lowest order non-linear couplings between three linear modes,
one unstable mode and two stable modes.
We assume OsC modes in the core for the unstable modes and $g$- and $r$-modes in the envelope for the stable modes.
OsC modes are rotationally stabilized convective modes and are strongly confined in the convective core when
the rotation rate of the core is small.
Assuming that the core of the stars rotates slightly faster than the envelope, 
\citet{LeeSaio20} and \citet{Lee2021} studied the modal properties of the OsC modes as a function of the rotation rate and
found that when the rotation rate is large enough the OsC modes resonantly excite $g$-modes to have
appreciable amplitudes in the envelope, that is, they have comparable amplitudes both in the core and in the envelope.
If OsC modes that resonantly excite envelope $g$-modes are common among rotating early type main sequence stars as suggested by \citet{Lee2021} for $M\gtrsim 2M_\odot$,
it is useful to estimate their amplitudes to compare with observations of low frequency variabilities
(e.g., \citet{Balona13, Balona17} for A-type stars, \citet{DegrooteAckeSamadietal2011,Balona16,Balonaetal19,BalonaOzuyar20} for B-type stars, and \citet{
VanReethMombargMathisetal2018} for $\gamma$ Dor stars) and to evaluate, for example, the efficiency of angular momentum transport by the low frequency modes in the rotating stars.
Non-linear coupling of the OsC modes with low frequency $g$- and $r$-modes in the envelope is 
a possible mechanism of amplitude saturation for the OsC modes.

To see the lowest order non-linear effects, 
we use three-mode coupling coefficients, which
we calculate in this paper employing the mathematical frame work described in \citet{Schenk_etal2001}.
Following this frame work, \citet{Lee2012} computed three-mode coupling coefficients using the traditional approximation of rotation (TAR) (e.g., \citealt{LeeSaio97}) for low frequency $g$- and $r$-modes in rotating SPB stars.
Although it is much easier to compute the low frequency modes of rotating stars using the TAR than the series expansion method for perturbations (e.g., \citealt{LeeSaio93}), we cannot resort to the TAR here, since it is not possible to properly calculate the eigenfunctions of the OsC modes in the core where buoyancy effects are negligible.
We thus use the series expansion method for the linear oscillation modes
of rotating stars.
In \S 2, we describe the method of solution we employ to derive the amplitude equations.
In \S 3 we discuss numerical results for amplitude evolutions.
\S 4 is for conclusion.

\section{Method of Solution}

\subsection{Linear Oscillation Modes in Differentially Rotating Stars}

The perturbed equation of motion for adiabatic oscillations of differentially rotating stars may be given by
(e.g., \citealt{LyndenBellOstriker1967})
\be
-\sigma^2\pmb{\xi}+\sigma\rmi\pmbmt{B}^\prime(\pmb{\xi})+\pmbmt{T}(\pmb{\xi})+\pmbmt{C}(\pmb{\xi})=0,
\label{eq:lo_eom}
\ee
where $\pmb{\xi}$ is the displacement vector, $\sigma$ is the oscillation frequency in the inertial frame, and the linear operators $\pmbmt{B}^\prime$, $\pmbmt{T}$,
and $\pmbmt{C}$ are defined as
\be
\pmbmt{B}^\prime(\pmb{\xi})
=2\Omega{\partial\over\partial\phi}\pmb{\xi}=2\left(\pmb{\Omega}\times\pmb{\xi}+\rmi m\Omega\pmb{\xi}\right),
\ee
\begin{align}
\pmbmt{T}(\pmb{\xi})&=\Omega^2{\partial^2\over\partial\phi^2}\pmb{\xi}\nonumber\\
&
=-m^2\Omega^2\pmb{\xi}
+2\rmi m\Omega\left(\pmb{\Omega}\times\pmb{\xi}\right)+\pmb{\Omega}\times\left(\pmb{\Omega}\times\pmb{\xi}\right),
\end{align}
\be
\pmbmt{C}(\pmb{\xi})=\delta\left({1\over\rho}\nabla p+\nabla\Phi\right),
\ee
where $\pmb{\Omega}$ is the vector of angular frequency of rotation of the star and $\Omega=|\pmb{\Omega}|$,
and $\rho$ and $p$ are the mass density and the pressure and $\Phi$ is the gravitational potential, and
$\delta$ indicates Lagrangian perturbation.
\citet{LyndenBellOstriker1967} showed that the operators $\rmi\pmbmt{B}^\prime(\pmb{\xi})$, 
$\pmbmt{T}(\pmb{\xi})$ and $\pmbmt{C}(\pmb{\xi})$ are Hermitian operators so that
\be
\left<\pmb{\eta},\pmbmt{O}(\pmb{\xi})\right>=\left<\pmb{\xi},\pmbmt{O}(\pmb{\eta})\right>^*,
\ee
where
\be
\left<\pmb{\eta},\pmbmt{O}(\pmb{\xi})\right>=\int \rho\pmb{\eta}^*\cdot\pmbmt{O}(\pmb{\xi})d^3\pmb{x},
\ee
and $\pmbmt{O}$ stands for each of the operators $\rmi\pmbmt{B}^\prime$, $\pmbmt{T}$,
and $\pmbmt{C}$, and
the asterisk $^*$ indicates complex conjugation.

Applying appropriate boundary conditions at the centre and surface of the stars, 
we solve equation (\ref{eq:lo_eom}) as an eigenvalue problem
to obtain
\be
-\sigma_A^2\pmb{\xi}_A+\sigma_A\rmi\pmbmt{B}^\prime(\pmb{\xi}_A)+\pmbmt{C}^\prime(\pmb{\xi}_A)=0,
\label{eq:eigeneq}
\ee
where $\pmbmt{C}^\prime=\pmbmt{T}+\pmbmt{C}$, and $\sigma_A$ and $\pmb{\xi}_A$ are the eigenfrequency and the eigenfunction and the subscript $A$ 
represents a collection of quantum numbers used to identify the oscillation modes.
Here, the boundary conditions we use are $\delta p=0$ at the surface and that $\pmb{\xi}$ is regular
at the stellar center.
When we assume $\pmb{\xi}\propto {\rm e}^{\rmi\sigma t}$, a mode with complex $\sigma$
is unstable (stable) if ${\rm Im}(\sigma)$ is negative (positive).
Multiplying by equation (\ref{eq:eigeneq}) an eigenfunction $\pmb{\xi}_B^*$ and integrating over the stellar volume, we obtain
\begin{align}
-\sigma_A^2\left<\pmb{\xi}_B,\pmb{\xi}_A\right>+\sigma_A\left<\pmb{\xi}_B,\rmi\pmbmt{B}^\prime(\pmb{\xi}_A)\right>
+\left<\pmb{\xi}_B,\pmbmt{C}^\prime(\pmb{\xi}_A)\right>=0.
\end{align}
Similarly, for an eigen-mode $(\sigma_B,\pmb{\xi}_B)$, we may obtain
\begin{align}
-(\sigma_B^*)^2\left<\pmb{\xi}_A,\pmb{\xi}_B\right>^*+\sigma_B^*\left<\pmb{\xi}_A,\rmi\pmbmt{B}^\prime(\pmb{\xi}_B)\right>^*
+\left<\pmb{\xi}_A,\pmbmt{C}^\prime(\pmb{\xi}_B)\right>^*=0.
\end{align}
Since 
$
\left<\pmb{\xi}_A,\pmb{\xi}_B\right>^*=\left<\pmb{\xi}_B,\pmb{\xi}_A\right>, 
$
and the operators $\rmi\pmbmt{B}^\prime$ and $\pmbmt{C}^\prime$ are Hermitian, 
we obtain 
\be
\left(\sigma_A-\sigma_B^*\right)\left[\left(\sigma_A+\sigma_B^*\right)\left<\pmb{\xi}_B,\pmb{\xi}_A\right>-\left<\pmb{\xi}_B,\rmi\pmbmt{B}^\prime(\pmb{\xi}_A)\right>\right]=0,
\ee
from which we may define a modified version of the orthogonal relation, applied for eigen-oscillations of rotating stars,
given by
\be
\left(\sigma_A+\sigma_B^*\right)\left<\pmb{\xi}_B,\pmb{\xi}_A\right>-\left<\pmb{\xi}_B,\rmi\pmbmt{B}^\prime(\pmb{\xi}_A)\right>
=\delta_{A,B}b_A.
\ee
where $\delta_{A,B}=1$ if $A=B$ and $\delta_{A,B}=0$ if $A\not=B$, and
\begin{align}
b_A&=2{\rm Re}(\sigma_A)\left<\pmb{\xi}_A,\pmb{\xi}_A\right>-\left<\pmb{\xi}_A,\rmi\pmbmt{B}^\prime(\pmb{\xi}_A)\right>
\nonumber\\
&=2\left<\pmb{\xi}_A,{\rm Re}(\omega_A)\pmb{\xi}_A\right>-\left<\pmb{\xi}_A,\rmi\pmbmt{B}(\pmb{\xi}_A)\right>,
\label{eq:ba_inertial}
\end{align}
where $\pmbmt{B}(\pmb{\xi})=2\pmb{\Omega}\times\pmb{\xi}$ and $\omega=\sigma+m\Omega$ is the frequency in the local
co-rotating frame and depends on $r$ for differential rotation $\Omega=\Omega(r)$.

In this paper, we compute oscillation modes of differentially rotating stars using
series expansions of the perturbations with finite expansion length. 
For example, the displacement vector $\pmb{\xi}$ may be
given in spherical polar coordinates by
\be
\pmb{\xi}=\xi^r\pmb{e}_r+\xi^\theta\pmb{e}_\theta+\xi^\phi\pmb{e}_\phi,
\ee
and the components $\xi^r$, $\xi^\theta$, and $\xi^\phi$ are expanded using spherical harmonic functions $Y_l^m(\theta,\phi)$ as
\be
{\xi^r\over r}=\sum_{j\ge1}S_{l_j}(r)Y_{l_j}^m{\rm e}^{\rmi \sigma t},
\label{eq:xir_expand}
\ee
\be
{\xi^\theta\over r}=\sum_{j\ge1}\left(H_{l_j}(r){\partial \over\partial\theta}Y_{l_j}^m
+T_{l'_j}(r){1\over\sin\theta}{\partial\over\partial\phi}Y_{l'_j}^m\right){\rm e}^{\rmi\sigma t},
\label{eq:xitheta_expand}
\ee
\be
{\xi^\phi\over r}=\sum_{j\ge1}\left(H_{l_j}(r){1\over\sin\theta}{\partial \over\partial\phi}Y_{l_j}^m
-T_{l'_j}(r){\partial\over\partial\theta}Y_{l'_j}^m\right){\rm e}^{\rmi\sigma t},
\label{eq:xiphi_expand}
\ee
and the pressure perturbation $p^\prime$, for example, may be given by
\be
p^\prime=\sum_{j\ge1}p^\prime_{l_j}(r)Y_{l_j}^m{\rm e}^{\rmi \sigma t},
\label{eq:p_expand}
\ee
where $Y_l^m=Y_l^m(\theta,\phi)$, and $l_j=|m|+2(j-1)$ and $l_j^\prime=l_j+1$ for even modes and
$l_j=|m|+2j-1$ and $l_j^\prime=l_j-1$ for odd modes where $j=1,~2,~\cdots,~j_{\rm max}$ and $j_{\rm max}$ may be
called the expansion length.
In this paper we use $j_{\rm max}=10$.
We confirm that the difference between the eigenfrequencies computed for $j_{\rm max}=10$ and for
$j_{\rm max}=8$, for example, are insignificant.
Note that the perturbation $p^\prime$ is symmetric about the equator of the star for even modes and
anti-symmetric for odd modes.
Substituting the expansions into the perturbed basic equations, we obtain a set of coupled linear ordinary 
differential equations for the expansion coefficients $S_l(r)$, $H_l(r)$, $p_l^\prime(r)$ and so on (see, e.g., \citealt{LeeSaio93}),
which may be solved with the boundary conditions at the centre and the surface of the star as
an eigenvalue problem for $\sigma$.
In general, the expansion coefficients $S_l$, $H_l$, $iT_{l'}$ and $p^\prime_l$ are real functions
for adiabatic modes having real eigenfrequency $\sigma$ but they can be complex if
the modes have complex eigenfrequency as expected for OsC modes.

\subsection{Amplitude Equations for Weakly Nonlinear Oscillations}

Non-linear evolution of small amplitude oscillation modes in rotating stars is governed by 
the oscillation equation with non-linear terms:
\be
\ddot{\pmb{\xi}}+\pmbmt{B}^\prime\left(\dot{\pmb{\xi}}\right)+\pmbmt{C}^\prime\left(\pmb{\xi}\right)=
\pmb{a}^{(2)}\left(\pmb{\xi},\pmb{\xi}\right),
\label{eq:nle}
\ee
where $\dot{\pmb{\xi}}=\partial\pmb{\xi}/\partial t$ and $\ddot{\pmb{\xi}}=\partial^2\pmb{\xi}/\partial t^2$, 
and $\pmb{a}^{(2)}\left(\pmb{\xi},\pmb{\xi}\right)$ represents a collection of nonlinear terms of second order in $\pmb{\xi}$, and the $i$ th component of $\pmb{a}^{(2)}$ is given by \citet{Schenk_etal2001}:
\begin{align}
a_i^{(2)}\left(\pmb{\xi},\pmb{\xi}\right)=&-{\rho^{-1}}\nabla_j\left\{p\left[\left(\Gamma_1-1\right)\Pi_i^j+\Xi_i^j
+\Psi\delta_i^j\right]\right\}\nonumber\\
&-({1/ 2})\xi^k\xi^l\nabla_k\nabla_l\nabla_i\Phi,
\end{align}
where $\nabla_j$ denotes the covariant derivative with respect to the coordinate $x^j$, 
\be
\Pi_i^j=(\nabla_i\xi^j)\nabla\cdot\pmb{\xi},
\ee
\be
\Xi_i^j=(\nabla_i\xi^k)(\nabla_k\xi^j), 
\ee
\be
\Psi={1\over 2}\Pi\left[\left(\Gamma_1-1\right)^2+{\partial\Gamma_1/\partial\ln\rho}\right]+{1\over 2}\left(\Gamma_1-1\right)\Xi,
\ee
\be
\Pi=\delta^i_j\Pi^j_i=\left(\nabla\cdot\pmb{\xi}\right)^2,  
\ee
\be
\Xi=\delta^i_j\Xi^j_i=\left(\nabla_j\xi^k\right)\left(\nabla_k\xi^j\right),
\ee
and $\delta^i_j$ is the Kronecker delta, 
and the repeated indices imply 
the summation from 1 to 3, and $\Gamma_1=\left(\partial\ln p/\partial\ln \rho\right)_{\rm ad}$.
Note that we have applied the Cowling approximation, neglecting $\Phi^\prime$, the Eulerian perturbation of 
the gravitational potential.

Following \citet{Schenk_etal2001}, we use eigenvalues $\sigma$ and eigenfunctions $\pmb{\xi}$ of the linear oscillation equation (\ref{eq:eigeneq}) to
expand the displacement vector $\pmb{\xi}(\pmb{x},t)$ and its time derivative $\dot{\pmb{\xi}}(\pmb{x},t)$ in the non-linear equation (\ref{eq:nle}):
\be
\begin{bmatrix}\pmb{\xi}(\pmb{x},t)\\
\dot{\pmb{\xi}}(\pmb{x},t)\end{bmatrix}
=\sum_Ac_A(t)\begin{bmatrix}
\pmb{\xi}_A(\pmb{x})\\ \rmi\sigma_A\pmb{\xi}_A(\pmb{x})\end{bmatrix},
\label{eq:nleexpand}
\ee
for which
\be 
\sum_A\left(\dot c_A-\rmi\sigma_A c_A\right)\pmb{\xi}_A(\pmb{x})=0.
\ee
Substituting the expansion (\ref{eq:nleexpand}) into the governing equation (\ref{eq:nle}),
making a scaler product with $\pmb{\xi}_A^*$ and integrating over the stellar volume, we obtain
\be
\dot c_A(t)-\rmi\sigma_Ac_A(t)=-{\rmi}\left<\pmb{\xi}_A,\pmb{a}^{(2)}\left(\pmb{\xi},\pmb{\xi}\right)\right>/b_A,
\ee
where
\be
c_A(t)=\left<\pmb{\xi}_A,\sigma_A^*\pmb{\xi}(t)-\rmi\dot{\pmb{\xi}}(t)-\rmi\pmbmt{B}\left(\pmb{\xi}(t)\right)\right>/b_A.
\ee
Substituting the expansion
$
\pmb{\xi}\left(\pmb{x},t\right)=\sum_Bc_B^*(t)
\pmb{\xi}_B^*\left(\pmb{x}\right)
$
into $\pmb{a}^{(2)}\left(\pmb{\xi},\pmb{\xi}\right)$, we obtain
\begin{align}
\dot c_A(t)+&\gamma_A c_A(t)-\rmi{\rm Re}(\sigma_A)c_A(t)\nonumber\\
&=-\rmi{\rm Re}(\sigma_A)s_A\sum_{B,C}
\eta^*_{ABC}c_B^*(t)c_C^*(t),
\label{eq:dotca}
\end{align}
where $\gamma_A={\rm Im}(\sigma_A)$, and 
\be
\eta_{ABC}=\eta(\pmb{\xi}_A,\pmb{\xi}_B,\pmb{\xi}_C)=\left<\pmb{\xi}_A^*,\pmb{a}^{(2)}\left(\pmb{\xi}_B,\pmb{\xi}_C\right)\right>/|\epsilon_A|, 
\ee
and $\epsilon_A\equiv {\rm Re}(\sigma_A) b_A$ and $s_A=\epsilon_A/|\epsilon_A|$.
Equation (\ref{eq:dotca}) may be considered the amplitude equations for weakly non-linear oscillations.
If we employ amplitude normalization of linear modes given by $|\epsilon_A|=GM^2/R$ where $M$ and $R$ denote the mass and the radius of the star and $G$
is the gravitational constant,
we obtain (e.g., \citealt{Schenk_etal2001})
\be
\eta_{ABC}=\eta_{ACB}=\eta_{BCA}=\eta_{BAC}=\eta_{CAB}=\eta_{CBA}.
\label{eq:kabc}
\ee
In equation (\ref{eq:dotca}), positive (negative) $\gamma_A$ represents linear stabilization (destabilization) of the mode $A$.
We also note that $\epsilon_A$ can be negative if ${\rm Re}(\sigma_A)$ and ${\rm Re}(\omega_A)$
have different signs.
If the second term on the right-hand-side of equation (\ref{eq:ba_inertial}) is negligible and uniform rotation
is a good approximation,
since ${\rm Re}(\sigma_A)>0$ (prograde) and ${\rm Re}(\omega_A)<0$ (retrograde) for $m<0$, for example, for $r$-modes,
we have $\epsilon_A<0$ for the modes.

For a mode $A$, let $\pmb{\xi}^0_A$ denote the displacement vector that satisfies the normalization condition $S_{l_1}(R)=1$ at the surface and let $\pmb{\xi}_A$ denote the displacement vector that satisfies $|\epsilon(\pmb{\xi}_A)|=GM^2/R$, and we write $\pmb{\xi}_A=f_A\pmb{\xi}_A^0$ using a scalar factor $f_A$.
Since $c_A^{0*}\pmb{\xi}_A^{0*}=c_A^*\pmb{\xi}_A^*$ and $|c_A^0\xi_A^{r0}(R)|=|f_Ac_A\xi_A^{r0}(R)|$, 
the surface amplitude of the radial component $\xi^r_A$ of the mode $A$
is approximately given by $|c_A^0|\sim|f_Ac_A|$ since $|\xi_A^{r0}(R)|\sim1$ because of the normalization $S_{l_1}(R)=1$.

To obtain non-zero coupling coefficient $\eta_{ABC}$, 
some selection rules between three oscillation modes have to be satisfied.
One of such selection rules is 
\be
m_A+m_B+m_C=0,
\label{eq:select_m}
\ee
for which we have
\be
{\rm Re}\left(\sigma_A+\sigma_B+\sigma_C\right)={\rm Re}\left(\omega_A+\omega_B+\omega_C\right).
\ee
Another selection rule may be simply stated that
the coupling coefficient $\eta_{ABC}$ is non-zero only when the mode triad consists of
three even modes or of one even mode and two odd modes (e.g., \citealt{Schenk_etal2001}).

\subsection{Amplitude Equations for Low Frequency Modes}

In this paper, we are interested in  parametric excitation and amplitude saturation of non-linearly coupled 
$g$-modes, $r$-modes, and OsC modes where
we may call the OsC modes parent modes and the stable $g$-modes and $r$-modes daughter modes.
We integrate the amplitude equations (\ref{eq:dotca}) to see evolutions of the mode amplitudes.

Let us rewrite the amplitude equations (\ref{eq:dotca}) into a more concrete form.
For this purpose, we use $a_i$, $b_j$, and $c_k$, instead of $c_A$, $c_B$, and $c_C$, to stand for
the modes and their amplitudes.
We let $a_1$ denote the complex amplitude of the parent mode and $b_i$ and $c_j$
those of the daughter modes and the modes $b_i$ and $c_j$ are assumed to belong to the mode sets $\pmb{y}_{m_b}$ and $\pmb{y}_{m_c}$ characterized by the azimuthal wave numbers $m_b$ and $m_c$, respectively.
If $m_b\not= m_c$, we may write
\be
\dot a_1=-\gamma_{a_1}a_1+\rmi{\rm Re}(\sigma_{a_1})a_1-2\rmi{\rm Re}(\sigma_{a_1})s^a_1\sum_{ij}\eta^*_{1ij}b^*_ic^*_j,
\label{eq:dotabc_a}
\ee
\be
\dot b_i=-\gamma_{b_i}b_i+\rmi{\rm Re}(\sigma_{b_i})b_i-2\rmi{\rm Re}(\sigma_{b_i})s_{b_i}\sum_j\eta^*_{i1j}a^*_1c^*_j,
\label{eq:dotabc_b}
\ee
\be
\dot c_j=-\gamma_{c_j}c_j+\rmi{\rm Re}(\sigma_{c_j})c_j-2\rmi{\rm Re}(\sigma{c_j})s^c_{j}\sum_i\eta^*_{j1i}a^*_1b^*_i,
\label{eq:dotabc_c}
\ee
and if $m_b= m_c$
\be
\dot a_1=-\gamma_{a_1}a_1+\rmi{\rm Re}(\sigma_{a_1})a_1-\rmi{\rm Re}(\sigma_{a_1})s_{a_1}\sum_{ij}\eta^*_{1ij}b^*_ib^*_j,
\label{eq:dotab_a}
\ee
\be
\dot b_i=-\gamma_{b_i}b_i+\rmi{\rm Re}(\sigma_{b_i})b_i-2\rmi{\rm Re}(\sigma_{b_i})s_{b_i}\sum_j\eta^*_{i1j}a^*_1b^*_j,
\label{eq:dotab_b}
\ee
where we have written for simplicity $\eta_{1ij}$ instead of $\eta_{a_1b_ic_j}$.
We thus obtain for $m_b\not=m_c$
\begin{align}
{d\over dt}&\left(|a_1|^2+\sum_j|b_j|^2+\sum_k|c_k|^2\right)=-2\gamma_{a_1}|a_1|^2-2\sum_j\gamma_{b_j}|b_j|^2\nonumber\\
&-2\sum_k\gamma_{c_k}|c_k|^2-4\sum_{j,k}\Delta\sigma_{1jk}^{abc}{\rm Im}(E_{1jk}^{abc}),
\end{align}
and for $m_b=m_c$
\begin{align}
{d\over dt}\left(|a_1|^2+\sum_j|b_j|^2\right)=&-2\gamma_{a_1}|a_1|^2-2\sum_j\gamma_{b_j}|b_j|^2
\nonumber\\
&-2\sum_{j,k}\Delta\sigma^{abb}_{1jj}{\rm Im}\left(E^{abb}_{1jk}\right),
\end{align}
where
\be
\Delta\sigma_{1jk}^{abc}={\rm Re}\left(\sigma_{a_1}s_{a_1}+\sigma_{b_j}s_{b_j}+\sigma_{c_k}s_{c_k}\right),
\ee
\be
E_{1jk}^{abc}=\eta_{1jk}a_1b_jc_k=\left|\eta_{1jk}a_1b_jc_k\right|
{\rm e}^{\rmi\varphi^{abc}_{1jk}}. 
\ee

Let us introduce the matrix $\pmbmt{K}^a_{bc}$ defined as
\be
\left(\pmbmt{K}^{a_i}_{bc}\right)_{jk}=\eta_{a_ib_jc_k}
\ee
for the parent mode $a_i$ and the two daughter modes $b_j$ and $c_k$.
For $m_b\not= m_c$, using $\pmbmt{K}^a_{bc}$, we may write the amplitude equation for
the parent modes $y_i^{m_a}=a_i$ for $i=1,~\cdots, ~n_a$ as
\begin{align}
\dot y_i^{m_a}=-g_{a_i}y_i^{m_a}
-2\rmi{\rm Re}(\sigma_{a_i})s_{a_i}\sum_{m_b,m_c}\left(\pmb{y}_{m_b}^{T}\pmbmt{K}_{bc}^{a_i}\pmb{y}_{m_c}\right)^*,
\label{eq:dy0a_mul}
\end{align}
and those for the daughter modes as
\be
\dot{\pmb{y}}_{m_b}=-\pmbmt{g}^b\pmb{y}_{m_b}-2\rmi\sum_i\left(y_i^{m_a}\pmbmt{f}^b\pmbmt{K}^{a_i}_{bc}\pmb{y}_{m_c}\right)^*,
\label{eq:dyb}
\ee
\be
\dot{\pmb{y}}_{m_c}=-\pmbmt{g}^c\pmb{y}_{m_c}-2\rmi\sum_i\left(y_i^{m_a}\pmbmt{f}^c(\pmbmt{K}^{a_i}_{bc})^T\pmb{y}_{m_b}\right)^*,
\label{eq:dyc}
\ee
where
$g_{a_i}=\gamma_{a_i}-\rmi{\rm Re}(\sigma_{a_i})$, and
\be
\pmb{y}_{m_b}=\left(\begin{array}{c} b_1\\ \vdots \\ b_{n_b}\\\end{array}\right),
\quad \pmb{y}_{m_c}=\left(\begin{array}{c} c_1\\ \vdots \\ c_{n_c}\\\end{array}\right),
\ee
\be
\pmbmt{f}^b=\left(\begin{array}{ccc}f_{b_1} & & 0 \\ & \ddots & \\ 0 &  & 
f_{b_{n_b}} \\\end{array}\right), 
\ee
\be
\pmbmt{g}^b=\left(\begin{array}{ccc} g_{b_1} & & 0\\
 &\ddots & \\
0 & & g_{b_{n_b}} \\
 \end{array}
 \right),
\ee
and $f_{b_j}={\rm Re}(\sigma_{b_j})s_{b_j}$ and $g_{b_j}=\gamma_{b_j}-\rmi{\rm Re}(\sigma_{b_j})$, and
$n_a$ is the number of the parent modes we consider and 
$n_b$ and $n_c$ denote the dimensions of the mode sets $\pmb{y}_{m_b}$ and $\pmb{y}_{m_c}$, respectively.
Note that the summation with respect to $m_b$ and $m_c$ in equation (\ref{eq:dy0a_mul}) must satisfy
the selection rule $m_a+m_b+m_c=0$, and also that we have implicitly assumed the summation in (\ref{eq:dy0a_mul}) includes
the summation over both even mode set pairs $(\pmb{y}_{m_b},\pmb{y}_{m_c})$ and odd mode set pairs.
For daughter modes we need to integrate the amplitude equations (\ref{eq:dyb}) and (\ref{eq:dyc})
for each of the combinations $(m_b,m_c)$.
For $m_b=m_c$, we may write the amplitude equations as
\be
\dot y_i^{m_a}=-g_{a_i}y_i^{m_a}
-\rmi{\rm Re}(\sigma_{a_i})s_{a_i}\sum\left(\pmb{y}_{m_b}^{T}\pmbmt{K}_{bb}^{a_i}\pmb{y}_{m_b}\right)^*,
\label{eq:dya0_b=c}
\ee
\be
\dot {\pmb{y}}_{m_b}=-\pmbmt{g}^{b}\pmb{y}_{m_b}-2\rmi \sum_i\left(y_i^{m_a}\pmbmt{f}^{b}\pmbmt{K}_{bb}^{a_i}\pmb{y}_{m_b}\right)^*,
\label{eq:dybc_b=c}
\ee
where the selection rule requires $m_a+2m_b=0$ and the summation in equation (\ref{eq:dya0_b=c}) indicates
the sum of both even mode set and odd mode set of $\pmb{y}_{m_b}$.
Note that we do not consider in this paper non-linear couplings between three stable linear modes $y_i^{m_a}$, 
$y_j^{m_b}$ and $y_k^{m_c}$.

When we consider OsC modes associated with different azimuthal wave numbers $m_a$ and $m^\prime_a$ 
for parent modes, there might appear
selection rules given by $m_a+m_b+m_c=0$ and $m_a^\prime+m_b^\prime+m_c=0$.
In this case, we have to rewrite the amplitude equations for $\pmb{y}_{m_c}$
as
\begin{align}
\dot{\pmb{y}}_{m_c}=-\pmbmt{g}^c\pmb{y}_{m_c}&-2\rmi\sum_i\left(y_i^{m_a}\pmbmt{f}^c(\pmbmt{K}^{a_i}_{bc})^T\pmb{y}_{m_b}\right)^*\nonumber\\
&-2\rmi\sum_i\left(y_i^{m^\prime_a}\pmbmt{f}^c(\pmbmt{K}^{a^\prime_i}_{b^\prime c})^T\pmb{y}_{m^\prime_b}\right)^*,
\end{align}
where $a_i^\prime$ indicates the $i$-th OsC modes of $m_a^\prime$.

\subsection{Adiabatic and Non-adiabatic Formulation of Non-linear Couplings}

In this paper we consider lowest order non-linear interactions between adiabatic oscillation modes
in rotating stars.
We use adiabatic eigenfrequencies and eigenfunctions to represent
non-linear oscillations that satisfy the adiabatic oscillation equations with non-linear terms (see equation
(\ref{eq:nle})).
Adiabatic oscillations are much easier to handle than non-adiabatic modes and 
adiabatic modes form a complete set of functions, which can be used to represent
the non-linear oscillations.
Adiabatic modes, however, ignore excitation and damping effects, which are important for amplitude evolutions governed by the amplitude equations (\ref{eq:dotca}).
So long as non-adiabatic effects are not significant, that is, $|\omega_{\rm I}/\omega_{\rm R}|\ll 1$,
one-to-one correspondence between
adiabatic modes and non-adiabatic modes is well established and the difference in $\sigma_{\rm R}$ between adiabatic modes and non-adiabatic modes is small.
This property is well satisfied for $g$-modes and $r$-modes unless the radial order is very high.
The growth rates and damping rates for adiabatic modes may be estimated by applying the quasi-adiabatic approximation or by simply calculating corresponding non-adiabatic modes (see \S 3.1.3 below).
This is the procedure we take in this paper.

To derive amplitude equations we may start with non-adiabatic oscillation equations as formulated by Buchler and colleagues (e.g., \citealt{BuchlerGoupil1984,BuchlerKovacs1986}).
They discussed the non-linear effects on radial pulsations of stars by using
the amplitude equations derived for non-adiabatic stellar pulsations.
Although Buchler and colleagues applied their formulation mainly to radial pulsators,
\citet{GoupilBuchler1994} later on extended the formalism for the amplitude equations to 
non-adiabatic and non-radial pulsations, assuming that
the linear non-adiabatic eigenmodes form a complete set of functions.
They derived the amplitude equations containing up to cubic terms in amplitude, which may
tend to suppress amplitude growth of unstable modes (see, also \citealt{BuchlerKovacs1986,KovacsBuchler1989}).
We expect that in the weak non-adiabatic limit, the adiabatic and non-adiabatic approaches to non-linear couplings lead to almost the same conclusions.
However, for non-adiabatic radial pulsations (e.g., \citealt{BuchlerGoupil1984}) and strange modes (e.g., \citealt{SaioJeffery1988}) fully non-adiabatic treatments of non-linear
mode interactions will be necessary to understand their amplitude evolution and saturation.

\section{Numerical Results}

We are interested in amplitude evolutions of low frequency oscillation modes driven by non-linear three-mode couplings to OsC modes in early type main sequence stars where the OsC modes are assumed to have 
appreciable amplitudes both in the core and in the envelope.

\subsection{Linear Mode Calculation}

As a background model for linear mode calculations, 
we employ a $2M_\odot$ zero age main sequence (ZAMS) star with the composition $X=0.7$ and $Z=0.02$.
The model has a convective core and an envelope, which is in radiative equilibrium except in
thin convective layers close to the stellar surface.
The model was computed by using a stellar evolution code originally written by \citet{Paczynski1970}.

To calculate low frequency modes in rotating stars,
we assume differential rotation $\Omega(r)$ given by
\be
\Omega(r)=\Omega_s\left[1+{p-1\over 1+e^{q(x-x_c)}}\right],
\label{eq:difrot}
\ee
where $x=r/R$, and $x_c$ denotes the outer boundary of the convective core, $\Omega_s$ is the rotation speed at the stellar surface, and $p$ and $q$ are parameters. 
Uniform rotation is given by $p=1$.
Setting $p>1$ implies that the core rotates faster than the envelope.
In this paper we use $q=100$, for which $\Omega(r)$ stays $\approx p\Omega_s$ for $x<x_c$ and $\approx \Omega_s$ for $x>x_c$
but changes steeply from $p\Omega_s$ to $\Omega_s$ at $x_c$. 
Since the envelope rotates almost uniformly,
we may use $\omega_s=\sigma+m\Omega(r=R)$
for the oscillation frequency observed in the co-rotating frame of the envelope.
No effects of rotation on the equilibrium structure are considered in this paper.

Some comments on the sign $s\equiv\epsilon/|\epsilon|$ may be appropriate.
We numerically find that $s$ is negative for OsC modes and
$r$-modes.
The negative $s_{\rm OsC}$ may be related to the fact that the OsC modes have negative energy of oscillation
as discussed by \citet{LeeSaio90} and \citet{Lee2021}.
The negative $s_{\rm r-modes}$, on the other hand, is related to the fact that the $r$-modes are 
observed retrograde in the local co-rotating frame in the envelope and prograde in the inertial frame, that is, the co-rotating frame frequency $\omega_s$ and the inertial frame frequency $\sigma$ have different signs.
For $g$-modes, $s_{\rm g-modes}$ is generally positive but can be negative when $\omega_s\sigma<0$.

It is important to note that for an eigenmode $(\sigma_m,\pmb{\xi}_m)$, we have
\be
{\rm Re}(\sigma_{-m})=-{\rm Re}(\sigma_m), \quad s_{-m}=s_m, 
\ee
so that ${\rm Re}(\sigma_{-m})s_{-m}=-{\rm Re}(\sigma_m)s_m$.
The product $s{\rm Re}(\sigma)$ has definite signs for the low frequency modes as tabulated in Table 1.

\begin{table}
\begin{center}
\caption{Sign of the product $s{\rm Re}(\sigma)$ for low frequency modes for $m<0$.
The signs are reversed for $m>0$.}
\begin{tabular}{ccc}
\hline
    & {\rm prograde} & {\rm retrograde}  \\
\hline
OsC-modes & $-$ & $\cdots$ \\
 g-modes   & $+$ & $-$  \\
 r-modes   & $\cdots$ & $-$  \\
\hline
\end{tabular}
\label{tab:ssig}
\end{center}
\end{table}

\subsubsection{Overstable Convective Modes}

We compute OsC modes assuming that the super-adiabatic temperature gradient in the convective core
has a finite small value $\nabla-\nabla_{\rm ad}=10^{-5}$ (\citealt{LeeSaio20,Lee2021}) where
$\nabla=d\ln T/d\ln p$ and $\nabla_{ad}=(\partial\ln T/\partial\ln p)_S$ with $S$ being the specific entropy.
In this paper, we assume slow rotation as given by $\overline\Omega_s\equiv\Omega_s/\sigma_0=0.2$ with
$\sigma_0=\sqrt{GM/R^3}=4.389\times10^{-4}{\rm rad/s}$ for the ZAMS model and weak
differential rotation given by $p=1.1$ and $q=100$.
The rotation law (\ref{eq:difrot}) for $p=1.1$ and $q=100$ corresponds to
a case where the core rotates slightly faster than the envelope although the core and the envelope themselves
rotate almost uniformly. 
Note that such a weak differential rotation law helps OsC modes to resonantly excite envelope $g$-modes (e.g., \citealt{Lee2021}).
The rotation law of this kind may be observationally anticipated for main sequence stars as summarized in \citet{AertsMathisRogers19} and \citet{Aerts2021}.
For the parameter values of $(\overline\Omega_s,p)$ we can rather easily calculate eigenmodes $(\sigma,\pmb{\xi})$ to prepare large sets of low frequency $g$-, $r$-modes and OsC modes 
when we use the series expansion method (\citealt{LeeSaio93}).
The traditional approximation of rotation (TAR) (e.g., \citealt{LeeSaio97}) may be used to compute $g$- and $r$-modes in the envelope
but the approximation is not necessarily appropriate for OsC modes in the core where the buoyant effect is vanishingly small.
We need correct eigenfunctions in the convective core to properly compute the coupling coefficients $\eta_{abc}$ between $g$- and $r$-modes and OsC modes.

In Table \ref{tab:osc}, we tabulate the eigenfrequency $\overline\sigma=\sigma/\sigma_0$ of the OsC modes of $m=-1$ and $m=-2$ at $\overline\Omega_s=0.2$ for $p=1.1$.
We use the same labelling for the OsC modes as that used in \cite{Lee19}, and the symbol
$B_n$ represents the OsC mode which has $n$ nodes of the eigenfunction in the convective core.
We find that the OsC modes $B_1$ and $B_2$ of $m=-1$ in the table have large amplitudes also in the envelope 
as a result of resonant excitation of the $g$-modes, and that
the $B_3$ mode is strongly confined in the convective core and does not have
any appreciable amplitudes in the envelope.
The same is true for the $m=-2$ OsC modes in the table, that is, although the OsC mode $B_1$ has appreciable amplitudes in the envelope, the $B_2$ mode does not.
At $\overline\Omega_s=0.2$ there exist two $m=-1$ OsC modes $B_1$ and $B_2$ which can be
parent modes to non-linearly excite $g$- and $r$-modes in the envelope. 
If the OsC modes are well confined into the core and have no appreciable amplitudes in the envelope, on the other hand, non-linear couplings of the OsC modes to the envelope modes will be very weak.

\begin{table}
\begin{center}
\caption{Complex eigen-frequency $\sigma=\sigma_{\rm R}+\rmi\gamma$ of adiabatic overstable convective modes of the $2M_\odot$ ZAMS model for $\overline\Omega_s=0.2$
where the notation $a(b)$ implies a number given by $a\times10^b$.
Normalized frequency $\overline\sigma=\sigma/\sqrt{GM/R^3}$ is tabulated.}
\begin{tabular}{cccc}
\hline
 $m$ & OsC mode & $\overline\sigma_{\rm R}$ & $\overline\gamma$ \\
 \hline
$-1$ &$B_1$   & 0.2206 & $-6.21(-5)$ \\
 $\cdots$ & $B_2$   & 0.2213 & $-1.71(-4)$  \\
$\cdots$ & $B_3$   & 0.2208 & $-1.42(-3)$  \\
$-2$ &$B_1$   & 0.4415 & $-1.51(-4)$ \\
$\cdots$ & $B_2$   & 0.4425 & $-2.14(-3)$ \\
\hline
\end{tabular}
\label{tab:osc}
\end{center}
\end{table}

\subsubsection{Low Frequency $g$- and $r$-Modes}

To calculate the coupling coefficients $\eta_{abc}$ for low frequency modes of a rotating star, 
we prepare the eigenfrequencies $\sigma$ and eigenfunctions $\pmb{\xi}$ for a large number of 
$g$- and $r$-modes.
For a given value of $m$, we calculate prograde and retrograde $g$-modes and retrograde $r$-modes, for
both even modes and odd modes.
Here, even (odd) modes have the pressure perturbation
$p^\prime(r,\theta,\phi)=\sum_{l}p_{l}^\prime(r) Y_{l}^m(\theta,\phi)$
which is symmetric (antisymmetric) with respect to the equator of the stars.
We consider only even (odd) modes whose dominant component $p_l^\prime Y_l^m$ in the expansion
tends to $p_{|m|}^\prime Y_{|m|}^m$ ($p_{|m|+1}^\prime Y_{|m|+1}^m$) in the limit of $\Omega_s\rightarrow 0$.
We compute the $g$-modes and $r$-modes for the radial order $n$ ranging from $n\sim 1$ to $n\sim 60$.
We prepare the mode sets for $|m|=1,~2$, and 3.

In this paper, we employ the OsC modes of $m_a=-1$ and $-2$ at $\overline\Omega_s=0.2$ for the parent modes
(see Table \ref{tab:osc}).
In the following, we use the subscript $a$ to denote parent modes and the subscripts $b$ and $c$ daughter modes.
The selection rule $m_a+m_b+m_c=0$ for non-zero coupling coefficients $\eta_{abc}$ is
a restriction to possible sets $(m_b,m_c)$ for a given $m_a$.
For the parent modes of $m_a=-1$, we consider $(m_b,m_c)=(-1,+2)$ and $(-2,+3)$  
while we consider $(m_b,m_c)=(-1,+3)$ and $(+1,+1)$ for the parent modes of $m_a=-2$.
Since the OsC modes belong to even modes, two daughter modes coupled to the OsC modes have to be both
even modes or both odd modes, which is the second selection rule used to restrict modes to be coupled.

We note that the mode sets we prepare for daughter modes occupy only a small fraction of the parameter space $(m_b,m_c,l_b,l_c,n_b,n_c)$ with $l_b\ge|m_b|$ and $l_c\ge|m_c|$.
We hope that the results obtained in this paper is of some use for further investigations of non-linear mode couplings of stellar non-radial pulsations.

\begin{figure}
\begin{center}
\mbox{\raisebox{3mm}{
\includegraphics[width=0.99\columnwidth]{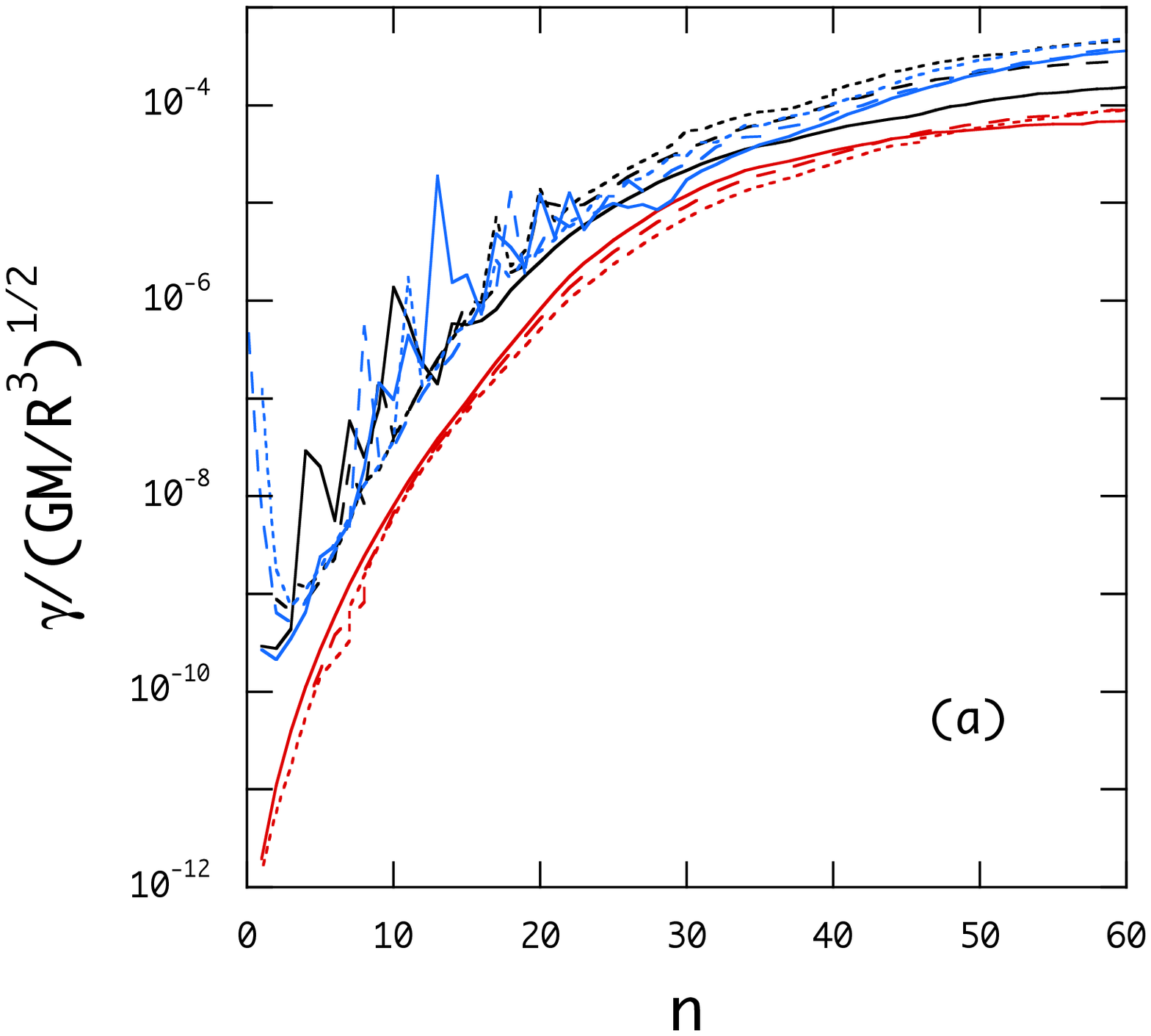}}
}

\mbox{\raisebox{0mm}{
\includegraphics[width=0.99\columnwidth]{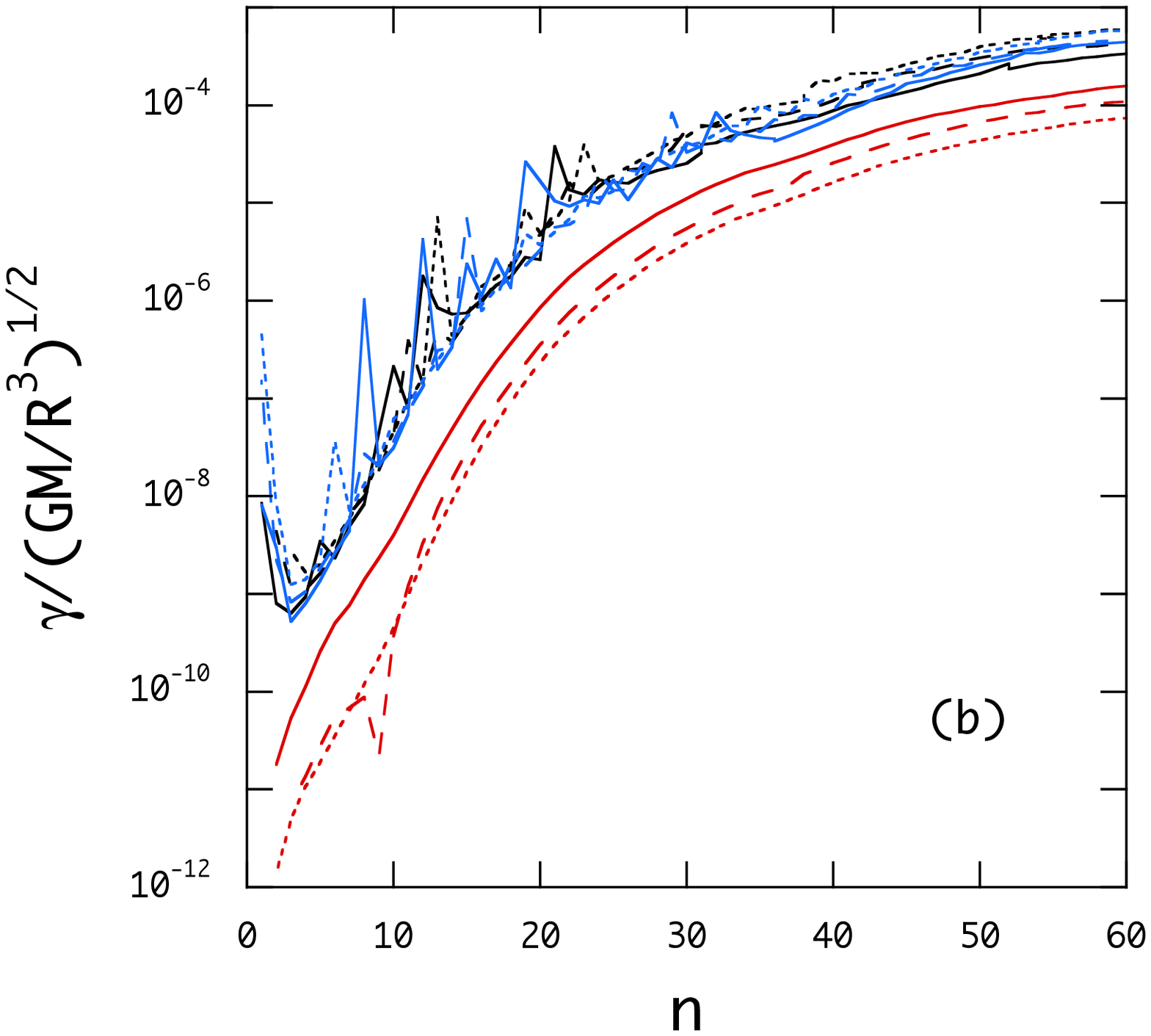}}
}
\caption{Damping rates $\gamma$ of low $|m|$ $g$- and $r$-modes versus the number $n$ of radial nodes
of the expansion coefficient $S_{l_1}$ for the $2M_\odot$ ZAMS model, where the black lines and blue lines
are respectively for the prograde and retrograde $g$-modes and the red lines are for the $r$-modes,
and the solid, dashed, and dotted lines are for $|m|=1$, 2, and 3, respectively.
Panels (a) and (b) are for even modes ($l=|m|$) and odd modes ($l=|m|+1$), respectively.
}
\label{fig:damprates}
\end{center}
\end{figure}

\begin{figure*}
\resizebox{0.66\columnwidth}{!}{
\includegraphics{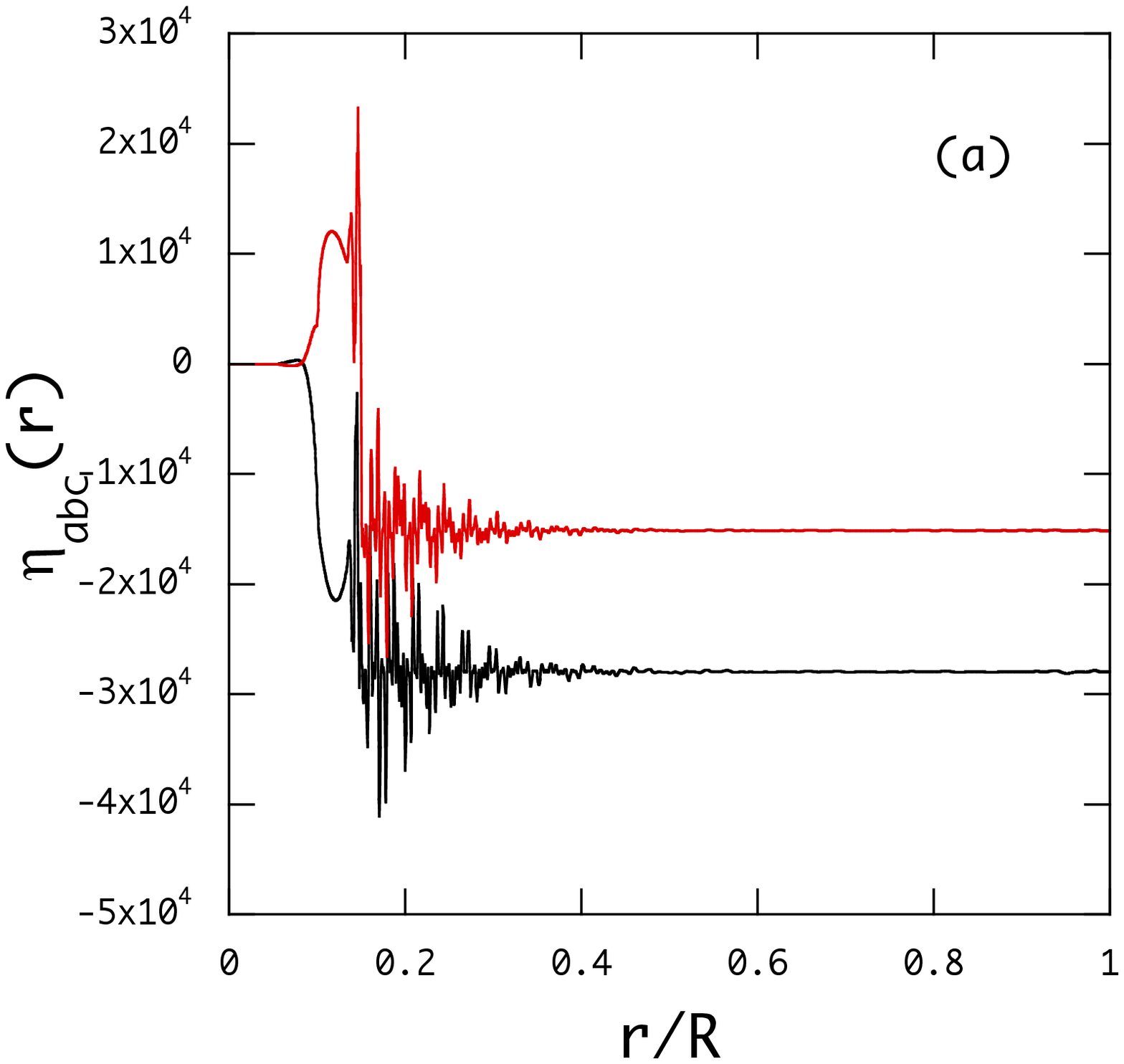}}
\resizebox{0.66\columnwidth}{!}{
\includegraphics{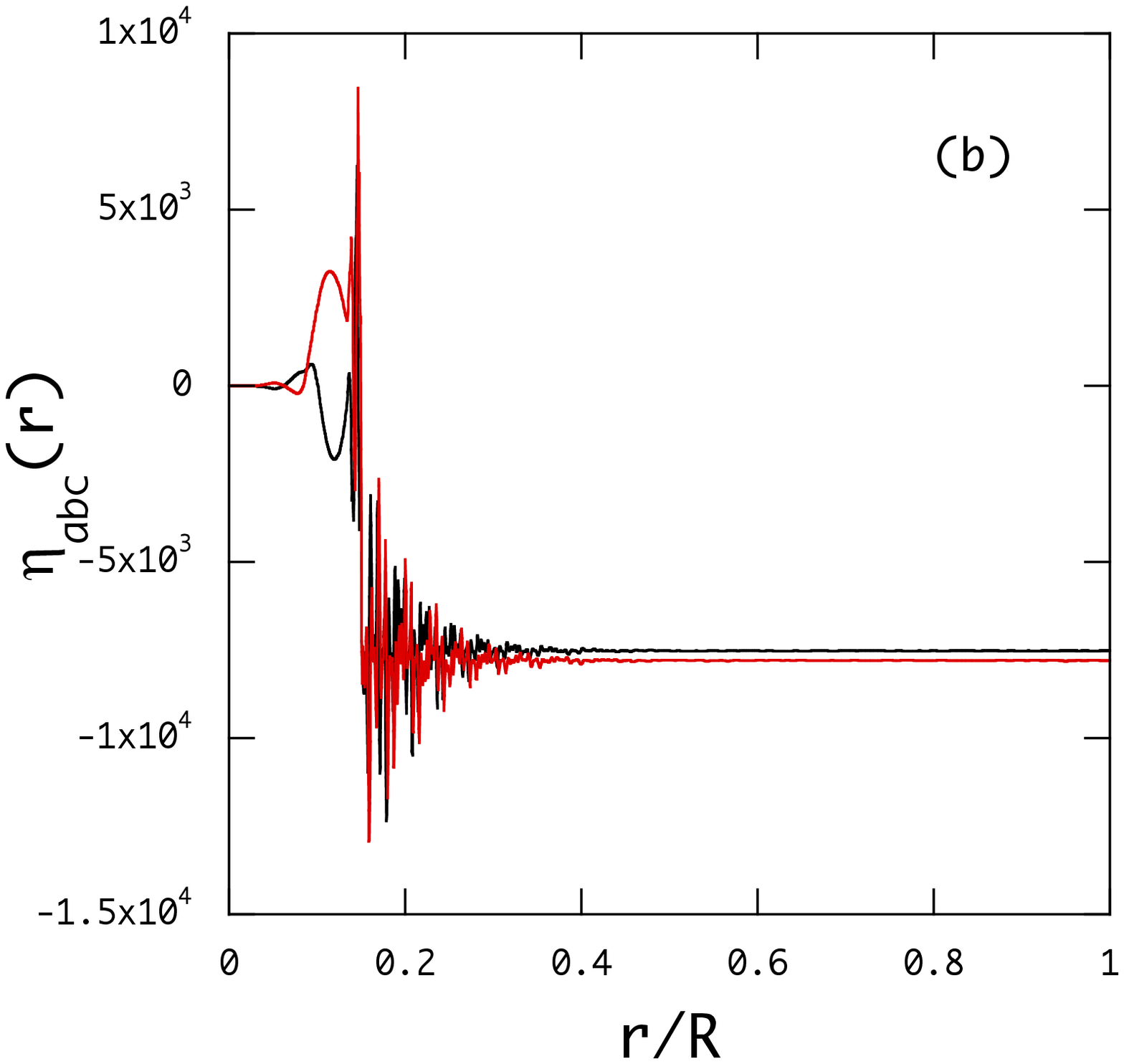}}
\resizebox{0.66\columnwidth}{!}{
\includegraphics{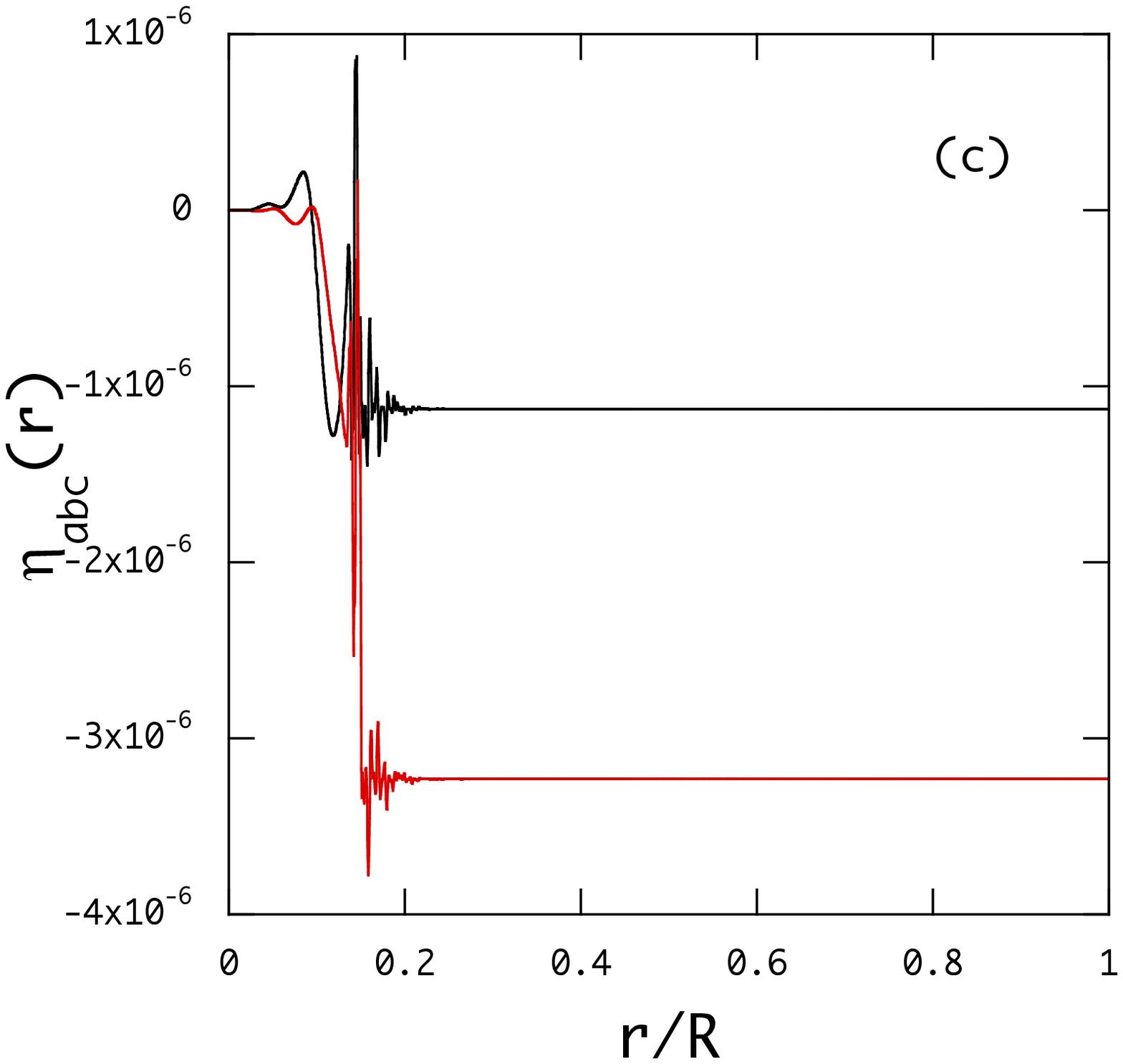}}
\caption{Complex $\eta_{abc}(r)$ as a function of $r/R$
for the $m=-1$ OsC mode $B_1$ (panel a), $B_2$ (panel b), and $B_3$ (panel c) as the parent mode 
where the daughter modes are odd $g_{40}^{-2}$- and $r_{50}^{+3}$-modes and
$\overline\Omega_s=0.2$.
Here, the black and red lines signify respectively the real part and imaginary part of the coefficient.
}
\label{fig:etaB123}
\end{figure*}

\subsubsection{Damping Rates}

Damping rates $\gamma$ of pulsation modes are important to determine amplitude evolutions of the modes.
We obtain damping rates of low frequency modes by carrying out non-adiabatic mode calculations 
for differentially rotating stars (e.g., \citealt{LeeSaio93,Lee2021}).
Non-adiabatic calculations of oscillation modes give the complex eigenfrequency $\sigma$ and we set $\gamma={\rm Im}(\sigma)$.
In Fig. \ref{fig:damprates}, we plot the damping rates $\overline\gamma=\gamma/\sqrt{GM/R^3}$
of low frequency $g$- and $r$-modes of the $2M_\odot$ ZAMS model versus the number $n$ of radial nodes of the expansion coefficient $S_{l_1}$ for $\overline\Omega_s=0.2$.
The damping rates $\overline\gamma$ for $n\sim1$ are much smaller than those for $n\sim 60$, for both $g$-modes
and $r$-modes.
In general the damping rates $\overline\gamma$ of $g$- and $r$-modes increase with increasing $n$, but
their rates of increase become small as $n$ increases.
When $n$ is small, however, $\overline\gamma$ of the $g$-modes of $l=|m|$, for example,
are occasionally disturbed by linear mode couplings with $g$-modes of $l=|m|+2,~|m|+4, \cdots$, 
which have much larger damping rates than the $l=|m|$ $g$-modes at a given frequency $\sigma$.
It is interesting to note that the damping rates of the $g$-modes for large $n$ are almost the same except for the even prograde $g$-modes of $|m|=1$.
The damping rates of $r$-modes are smaller by a factor $\sim 5$ than those of the $g$-modes.
The figure also suggests that $\overline\gamma$ of the even $r$-modes only weakly depends on $|m|$,
but $\overline\gamma$ of the odd $r$-modes decreases as $|m|$ increases.

\subsubsection{Coupling Coefficient $\eta_{abc}$}

\begin{figure}
\resizebox{1\columnwidth}{!}{
\includegraphics{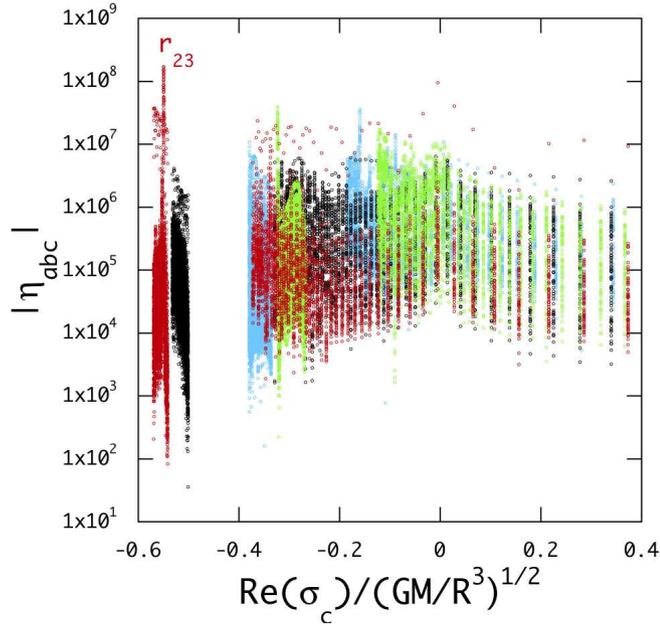}}
\caption{$|\eta_{abc}|$ versus ${\rm Re}(\overline\sigma_{c})$ for the couplings of the $m_a=-1$ OsC mode $B_1$ to $g$- and $r$-modes of $m_b=-1$ and $-2$ and 
of $m_c=+2$ and $+3$.
The black (red) dots represent $|\eta_{abc}|$ for the couplings to the odd (even) daughter modes with $(m_b,m_c)=(-2,+3)$ while the green (light blue) dots for the couplings to the odd (even) modes with $(m_b,m_c)=(-1,+2)$.
}
\label{fig:eta_mm1}
\end{figure}

We use the formulation given by \citet{Schenk_etal2001} to compute the coupling coefficient $\eta_{abc}$ for low frequency modes of rotating stars where
the eigenfrequencies $\sigma$ and eigenfunctions $\pmb{\xi}$ are those of adiabatic modes.

If we define the function $\eta_{abc}(r)$ as
\be
\eta_{abc}(r)={R\over GM^2}\int_0^r \overline{\pmb{\xi}_a\cdot\pmb{a}^{(2)}\left(\pmb{\xi}_b,\pmb{\xi}_c\right)}\rho r^2 dr,
\ee
where
\be
\overline{f}=\int_0^{2\pi}\int_0^\pi f\sin\theta d\theta d\phi,
\ee
we have the coupling coefficient $\eta_{abc}=\eta_{abc}(R)$.
In Fig.\ref{fig:etaB123}, we plot $\eta_{abc}(r)$ versus $r/R$ for the couplings between
the $m=-1$ OsC modes and $g_{40}^{-2}$- and $r_{50}^{+3}$-modes where the $g$- and $r$-modes belong to odd modes.
Note that because the OsC modes have complex $\sigma$ and $\pmb{\xi}$,
the coupling coefficient $\eta_{abc}$ becomes complex even if the adiabatic $g$- and $r$-modes
have real eigenfunctions and eigenfrequencies.
Since the low frequency modes we use have many nodes of the eigenfunctions in the envelope,
$\eta_{abc}(r)$ shows rapid fluctuations around a mean $\left<{\eta_{abc}(r)}\right>$ as a function of $r$ 
although the amplitudes of the fluctuations quickly 
damp toward the stellar surface.
The mean values $\left<\eta_{abc}(r)\right>$
in the envelope may be those attained at the interface between the convective core and the radiative envelope.
Since the OsC $B_3$ mode is strongly confined into the core and practically have
no amplitudes in the envelope, its coupling with the $g$- and $r$-modes in the envelope 
is very weak and hence the coupling coefficients is very small.

In Fig. \ref{fig:eta_mm1}, for the $m_a=-1$ OsC mode $B_1$
we plot the coupling coefficient $|\eta_{abc}|$ versus
$\overline\sigma_{c\rm R}={\rm Re}(\overline\sigma_c)$ of the daughter modes in the sets $\pmb{y}_{m_c}$ for $m_c=+2$ and $+3$, for which the other mode sets are given by $\pmb{y}_{m_b}$ for $m_b=-1$ and $-2$.
For a given combination $(m_b,m_c)$, mode sets $\pmb{y}_{m_b}$ and $\pmb{y}_{m_c}$ are both
composed of even modes or of odd modes.
Note that $|\eta_{abc}|$'s for different $\overline\sigma_{b{\rm R}}$'s appear
as vertically aligned dots at a given $\overline\sigma_{c{\rm R}}$.
In the figure, the red dots, for example, represent $|\eta_{abc}|$ for the even mode set $\pmb{y}_{m_c=+3}$, 
and they are separated into two groups of dots, 
one for the $g$-modes and the other for the $r$-modes.
The gap between the two mode groups appears because we have considered only a limited set of the $g$-modes with the radial order $n\lesssim 60$.
For a given $m$, the frequency $\sigma_{\rm R}$ of the lowest radial order $r$-mode
is approximately given by
\be
\sigma_{\rm R}\approx-{(l'+2)(l'-1)\over l'(l'+1)}m\Omega_e,
\ee
where $\Omega_e$ is the rotation frequency in the envelope and is approximately equal to that
at the stellar surface, and $l'=|m|+1$ for even modes and $l'=|m|$ for odd modes.
Note that $|\sigma_{\rm R}|$ for $r$-modes increases as the radial order increases.
For $m=1$, $2$, and $3$, we respectively obtain $\sigma_{\rm R}/\Omega_s\approx -2/3$, $-5/3$, and $-27/10$ for even modes, and $\sigma_{\rm R}/\Omega_s\approx 0$, $-4/3$, and $-5/2$ for odd modes.
For $\overline\Omega_s=0.2$, the lowest radial order $r$-modes of $m=3$, for example, appear at 
$\overline\sigma_{\rm R}\approx -0.5$ for odd modes, and at $\overline\sigma_{\rm R}\approx -0.54$
for even modes.

\begin{figure}
\resizebox{0.49\columnwidth}{!}{
\includegraphics{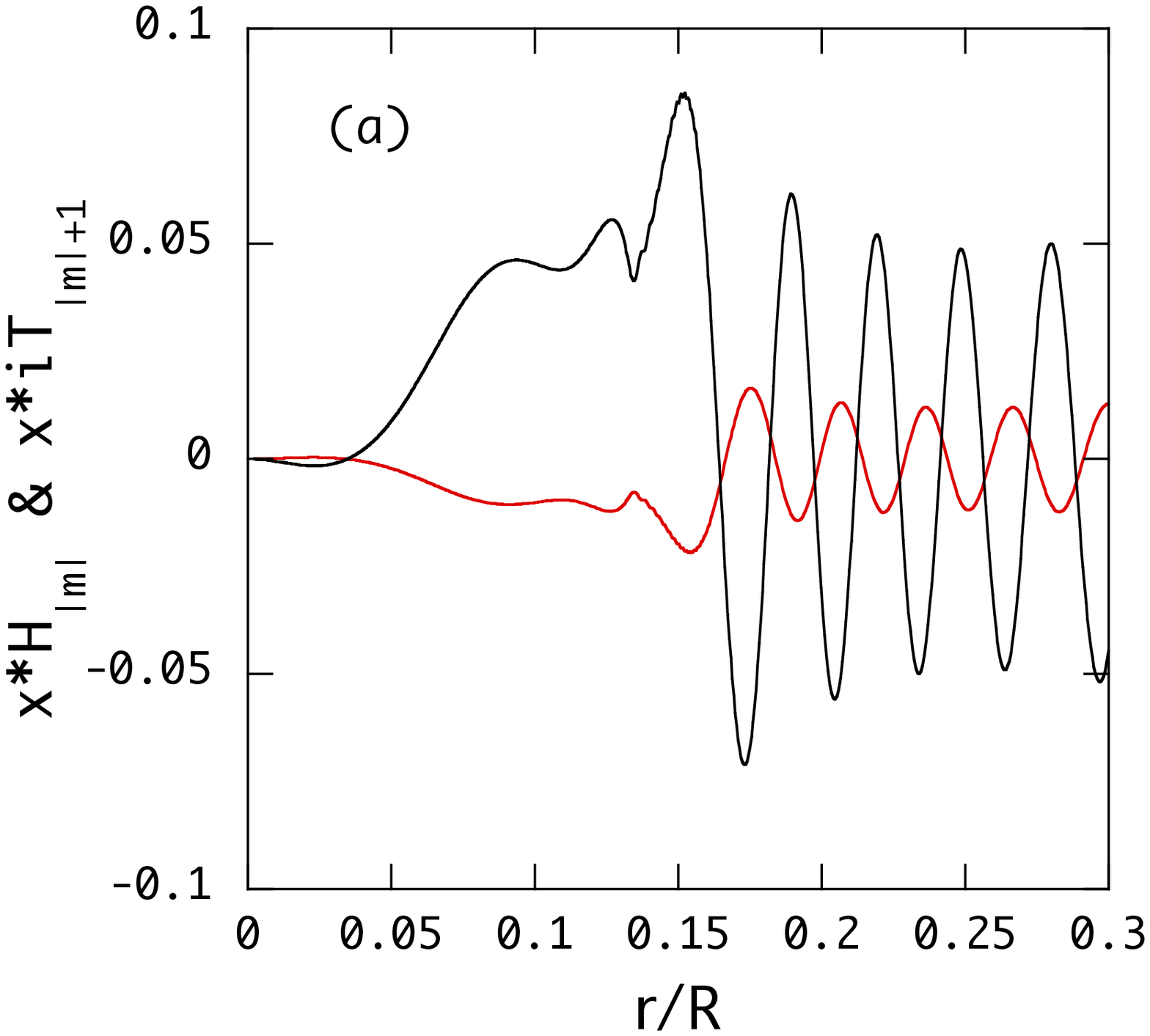}}
\resizebox{0.49\columnwidth}{!}{
\includegraphics{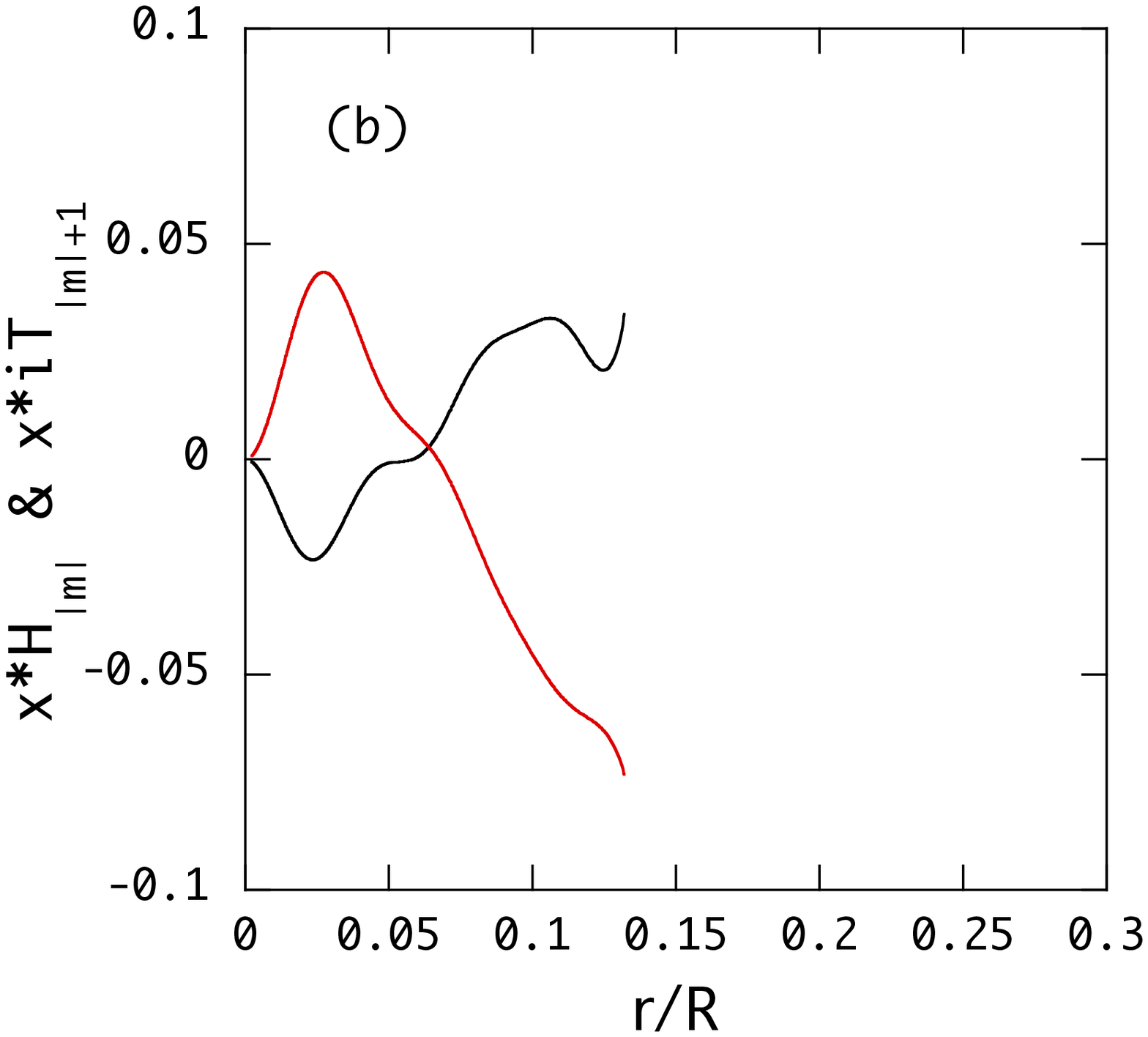}}
\caption{Expansion coefficients $xH_{|m|}$ (black lines) and $x\rmi T_{|m|+1}$ (red lines) versus $x=r/R$ for the $m=3$ even $r_{23}$-mode (panel a) and for an $m=3$ core mode, which is obtained by imposing the outer boundary condition $d(p^\prime/\rho g r)/dr=0$ at the interface between the convective core and the envelope (panel b).
}
\label{fig:rmoderesonance}
\end{figure}

The largest value of $|\eta_{abc}|$ is only weakly dependent on $\overline\sigma_{c\rm R}$
and in general less than $\sim 10^8$ although there exist a few exceptions.
In Fig.\ref{fig:eta_mm1}, we find that $|\eta_{abc}|$ exceeds $\sim 10^8$ at ${\rm Re}(\overline\sigma_c)\approx-0.5501$, corresponding to the even $r_{23}$-mode of $m_c=+3$, and that
$|\eta_{abc}|$ at this ${\rm Re}(\overline\sigma_c)$ is largest for the $m_b=-2$ $g_{58}$-mode.
The reason for the large $|\eta_{abc}|$'s may be attributed to the fact that the $r_{23}$-mode is in resonance with an inertial mode in the core.
Fig.\ref{fig:rmoderesonance} plots
the eigenfunctions $xH_{|m|}$ and $x\rmi T_{|m|+1}$ versus
$x=r/R$ for the $r_{23}$-mode (panel a) and those of
a core mode of the frequency $\overline\sigma_{\rm R}\approx -0.550$ which is obtained by imposing the boundary condition $d(p^\prime/\rho g r)/dr=0$ at the interface at $r_c/R\approx 0.132$ (panel b).
We find that the behavior of the eigenfunctions against $r/R$ in the core is quite similar
in the sense that the numbers of radial nodes in the core are the same for the two modes.
Since the magnitudes of $|\eta_{abc}|$ are approximately equal to those of $|\eta_{abc}(r)|$ attained 
near the interface between the core and the envelope, the resonant enhancement of the amplitudes of the
$r_{23}$-mode around the interface leads to the enhancement of $|\eta_{abc}|$.
We also find that similar resonance phenomena happen between $g$-modes around the $g^{-2}_{58}$-mode in the envelope and an inertial mode in the core.
It is interesting to note that \citet{Ouazzanietal20} and \citet{Saioetal21} have recently discussed resonances between $g$-modes in the envelope and inertial modes in the core 
to explain the dips observationally found in the period - period difference relations for $\gamma$ Dor stars.

\subsection{Three-Mode Coupling}

For non-linear three-mode coupling between modes $(\sigma_a,\pmb{\xi}_a)$, $(\sigma_b,\pmb{\xi}_b)$, and $(\sigma_c,\pmb{\xi}_c)$,
we obtain, for example, from equations (\ref{eq:dotabc_a}) to (\ref{eq:dotabc_c})  
\begin{align}
\dot a+\gamma_a a-\rmi{\rm Re}(\sigma_a)a
=-2\rmi{\rm Re}(\sigma_a)s_a\eta^*_{abc} b^*c^*,
\label{eq:dotc1}
\end{align}
\begin{align}
\dot b+\gamma_b b-\rmi{\rm Re}(\sigma_b)b
=-2\rmi{\rm Re}(\sigma_b)s_b\eta^*_{abc} c^*a^*,
\label{eq:dotc2}
\end{align}
\begin{align}
\dot c+\gamma_c c-\rmi{\rm Re}(\sigma_c)c
=-2\rmi{\rm Re}(\sigma_c)s_c\eta^*_{abc} a^*b^*.
\label{eq:dotc3}
\end{align}
Here, we regard the mode $a$ as the unstable parent mode and the modes $b$ and $c$ as stable daughter modes.
It is interesting to note that introducing the dependent variables $\tilde a$ defined by 
$\tilde a=|\eta_{abc}|{\rm e}^{-\rmi{\rm Re}(\sigma_a)t}a$ with $\eta_{abc}=|\eta_{abc}|e^{\rmi\delta_{abc}}$,
equation (\ref{eq:dotc1}), for example, reduces to
\be
\dot{\tilde a}=-\gamma_a\tilde a-2\rmi{\rm Re}(\sigma_a)s_a\tilde b^*\tilde c^*{\rm e}^{-\rmi\Delta\sigma_{abc}-\rmi\delta_{abc}},
\label{eq:no-eta}
\ee
where  
$
\Delta\sigma_{abc}={\rm Re}(\sigma_a+\sigma_b+\sigma_c).
$
Equation (\ref{eq:no-eta}) shows that the coupling coefficient $|\eta_{abc}|$ becomes implicit in the three-mode coupling equation and
suggests that the growth and damping rates $\gamma_j$ play an essential role
to determine the qualitative property of the amplitude evolution.

Substituting $a=\hat ae^{\rmi \varphi_a}$ so on into equations (\ref{eq:dotc1}) to (\ref{eq:dotc3})
where $\hat a=|a|$ and $\varphi_a$ being real quantities,
we obtain
\be
\dot {\hat a}=-\gamma_a{\hat a}-2{\rm Re}(\sigma_a)s_a|\eta_{abc}|\hat b\hat c\sin\varphi,
\label{eq:dothata}
\ee
\be
\dot {\hat b}=-\gamma_b{\hat b}-2{\rm Re}(\sigma_b)s_b|\eta_{abc}|\hat c\hat a\sin\varphi,
\label{eq:dothatb}
\ee
\be
\dot {\hat c}=-\gamma_c{\hat c}-2{\rm Re}(\sigma_c)s_c|\eta_{abc}|\hat a\hat b\sin\varphi,
\label{eq:dothatc}
\ee
\begin{align}
\dot\varphi=\Delta\sigma_{abc}-2|\eta_{abc}|\cos\varphi &\left[{{\rm Re}(\sigma_a)s_a\hat b\hat c\over \hat a} +{{\rm Re}(\sigma_b)s_b\hat c\hat a\over \hat b}\right. \nonumber\\
& \left.
\quad +{{\rm Re}(\sigma_c)s_c\hat a\hat b\over \hat c} \right],
\end{align}
where 
\be
\varphi=\varphi_a+\varphi_b+\varphi_c+\delta_{abc}.
\ee
This form of the amplitude equations can be used to derive the conditions for parametric instability and 
for the stability of equilibrium states of the amplitudes.

Parametric instability occurs between one unstable ($\gamma_a<0$) parent mode and two stable ($\gamma_b>0$ and
$\gamma_c>0$) daughter modes when the amplitude of the parent mode, $|a|$, exceeds
the critical amplitude $|a_{\rm crit}|$ given by (e.g., \citealt{Dziembowski1982,Arras_etal2003})
\be
|a_{\rm crit}|^2={1\over 4|\eta_{abc}|^2Q_bQ_c}\left[1+\left({\Delta\sigma_{abc}\over\gamma_b+\gamma_c}\right)^2\right],
\label{eq:critampl}
\ee
where $Q_j={\rm Re}(\sigma_j)s_j/\gamma_j$.

Similarly, the equilibrium amplitude of the parent mode is given by (e.g., \citealt{Arras_etal2003})
\be
|a_{e}|^2={1\over4|\eta_{abc}|^2Q_bQ_c}\left[1+\left({\Delta \sigma_{abc}\over\Delta\gamma_{abc}}\right)^2\right],
\ee
with which those for the daughter modes are given by
\be
\left|{b_{e}}\right|^2=\left|a_{e}\right|^2{Q_b/ Q_a}, \quad
\left|{c_{e}}\right|^2=\left|a_{e}\right|^2{Q_c/ Q_a},
\ee
where $\Delta\gamma_{abc}=\gamma_a+\gamma_b+\gamma_c$.
Here, we have assumed $Q_bQ_c>0$, $Q_cQ_a>0$, and $Q_aQ_b>0$, that is, 
$Q_a$, $Q_b$, and $Q_c$ all have the same sign.
For the parametric instability, since
the parent mode $a$ is assumed unstable ($\gamma_a<0$) and the daughter modes $b$ and $c$
stable ($\gamma_b>0$ and $\gamma_c>0$), the sign of ${\rm Re}(\sigma_a)s_a$ is different from those of
${\rm Re}(\sigma_b)s_b$ and ${\rm Re}(\sigma_c)s_c$, which have the same sign.
We may consider $\Delta\gamma_{abc}>0$ as the criteria for effectively stable equilibrium states 
produced by three-mode couplings (e.g., \citealt{WuGoldreich2001,Arras_etal2003}).

\begin{figure*}
\resizebox{0.5\columnwidth}{!}{
\includegraphics{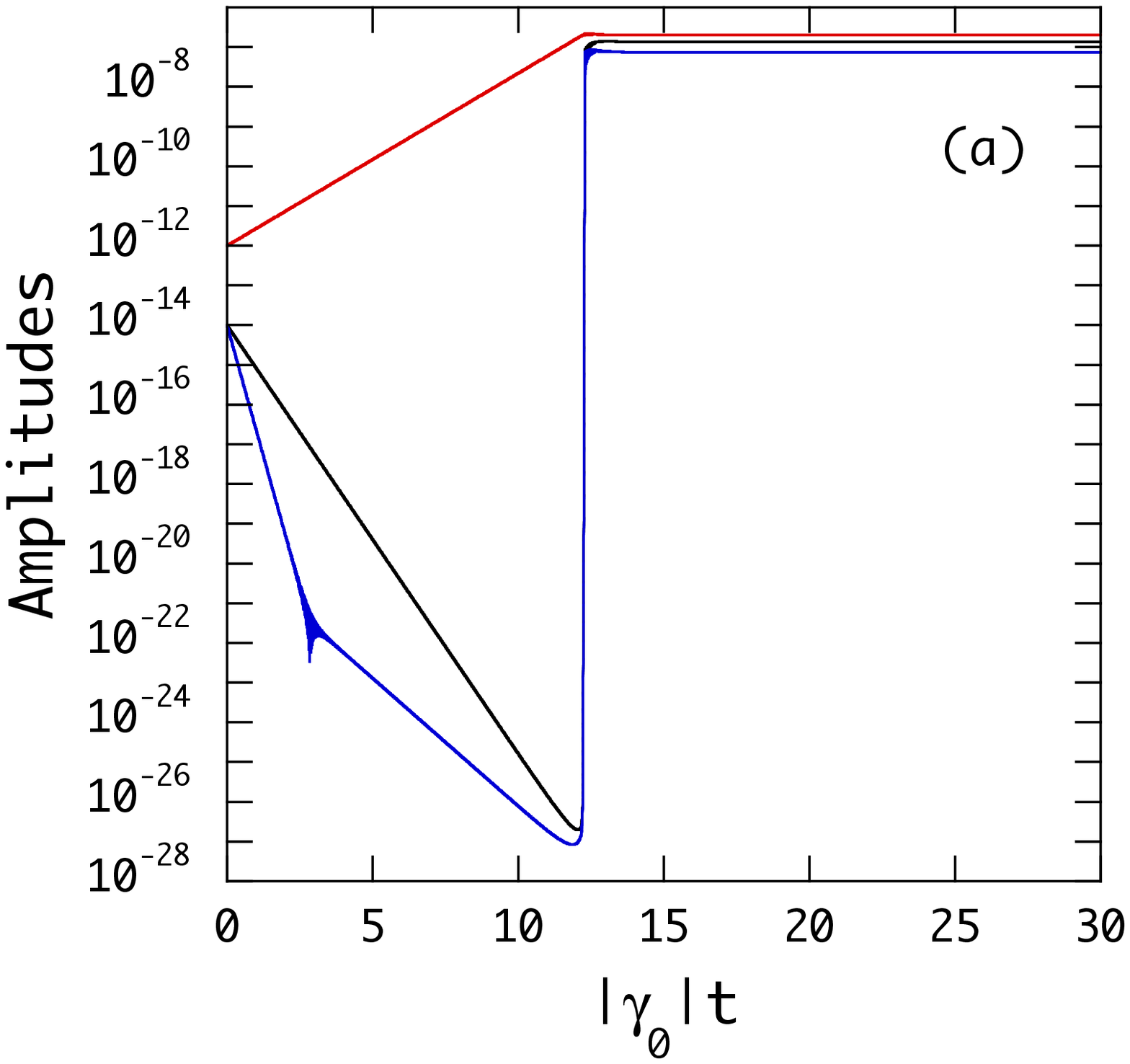}}
\resizebox{0.5\columnwidth}{!}{
\includegraphics{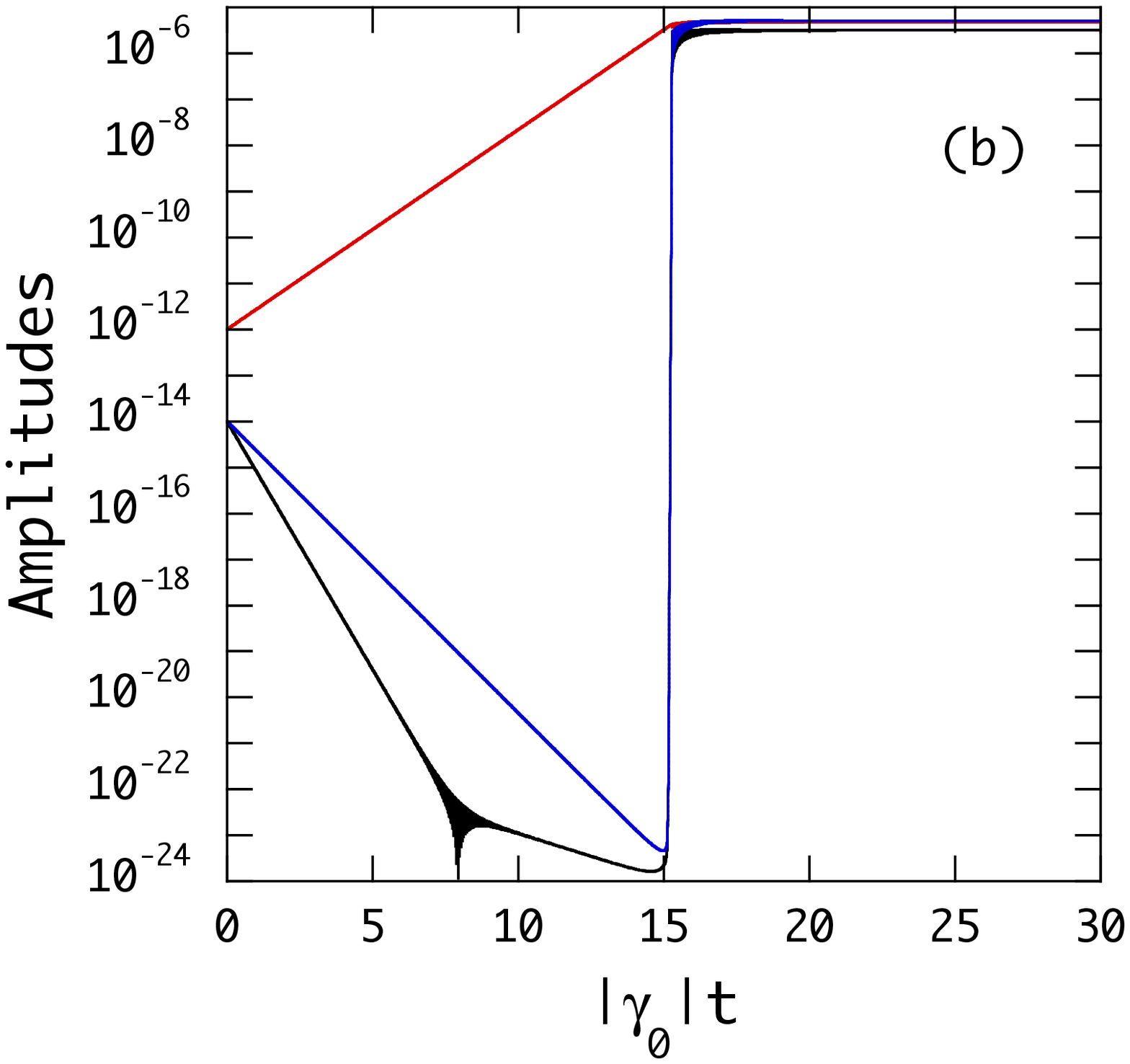}}
\resizebox{0.5\columnwidth}{!}{
\includegraphics{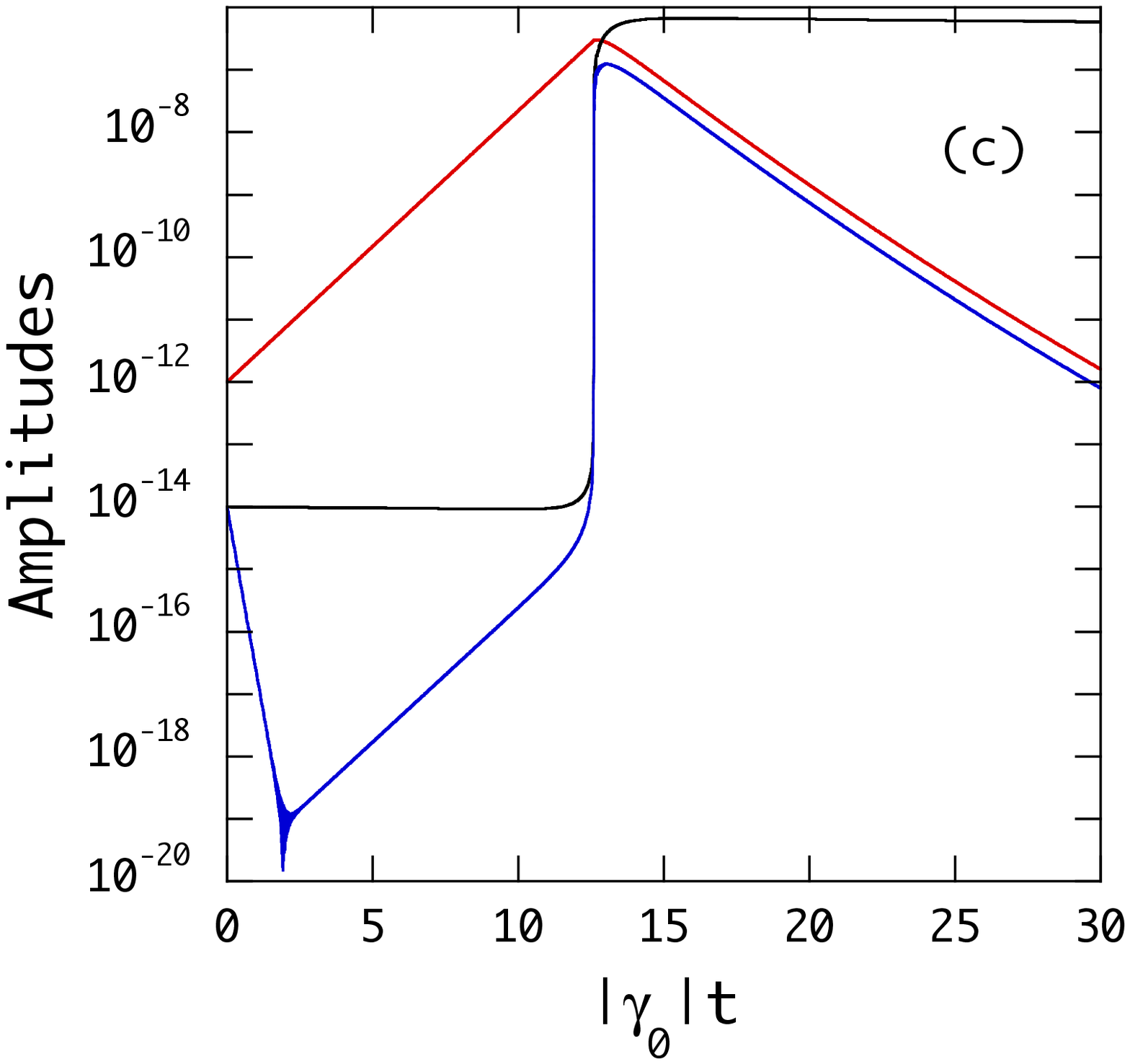}}
\resizebox{0.5\columnwidth}{!}{
\includegraphics{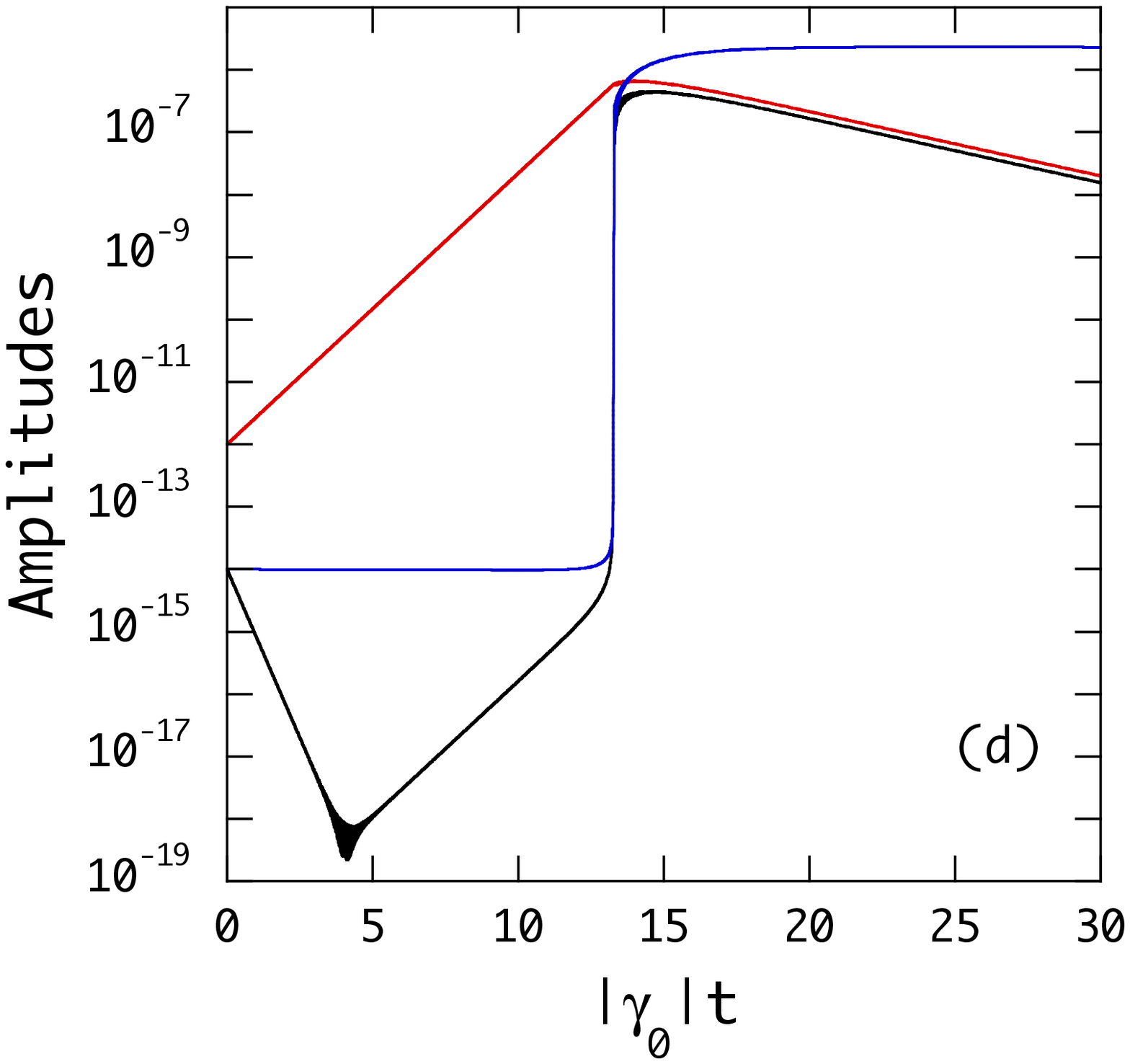}}
\caption{Amplitude evolution of the low frequency modes of $m_b=-1$ and $m_c=+2$ excited by the $m_a=-1$ OsC mode $B_1$ where $|\overline\gamma_0|=6.21\times10^{-5}$ and $\overline\Omega_s=0.2$.
The panels (a) to (d) respectively shows the evolutions of the mode pairs $(g_{60}^{-1},g_{60}^{+2})$, $(g_{60}^{-1},r_{60}^{+2})$, $(g_{15}^{-1},g_{60}^{+2})$, and $(g_{60}^{-1},r_{15}^{+2})$ coupled to the OsC mode $B_1$ where the notations $g_n^m$ and $r_n^m$ stand for the $g$-modes and $r$-modes of the radial order $n$ and azimuthal wavenumber $m$.
The red lines stand for the $m_a=-1$ $B_1$ OsC mode and the black and blue lines for the low frequency modes with $m_b=-1$ and $m_c=+2$, respectively.}
\label{fig:amp_thmc}
\end{figure*}

\begin{figure*}
\resizebox{0.5\columnwidth}{!}{
\includegraphics{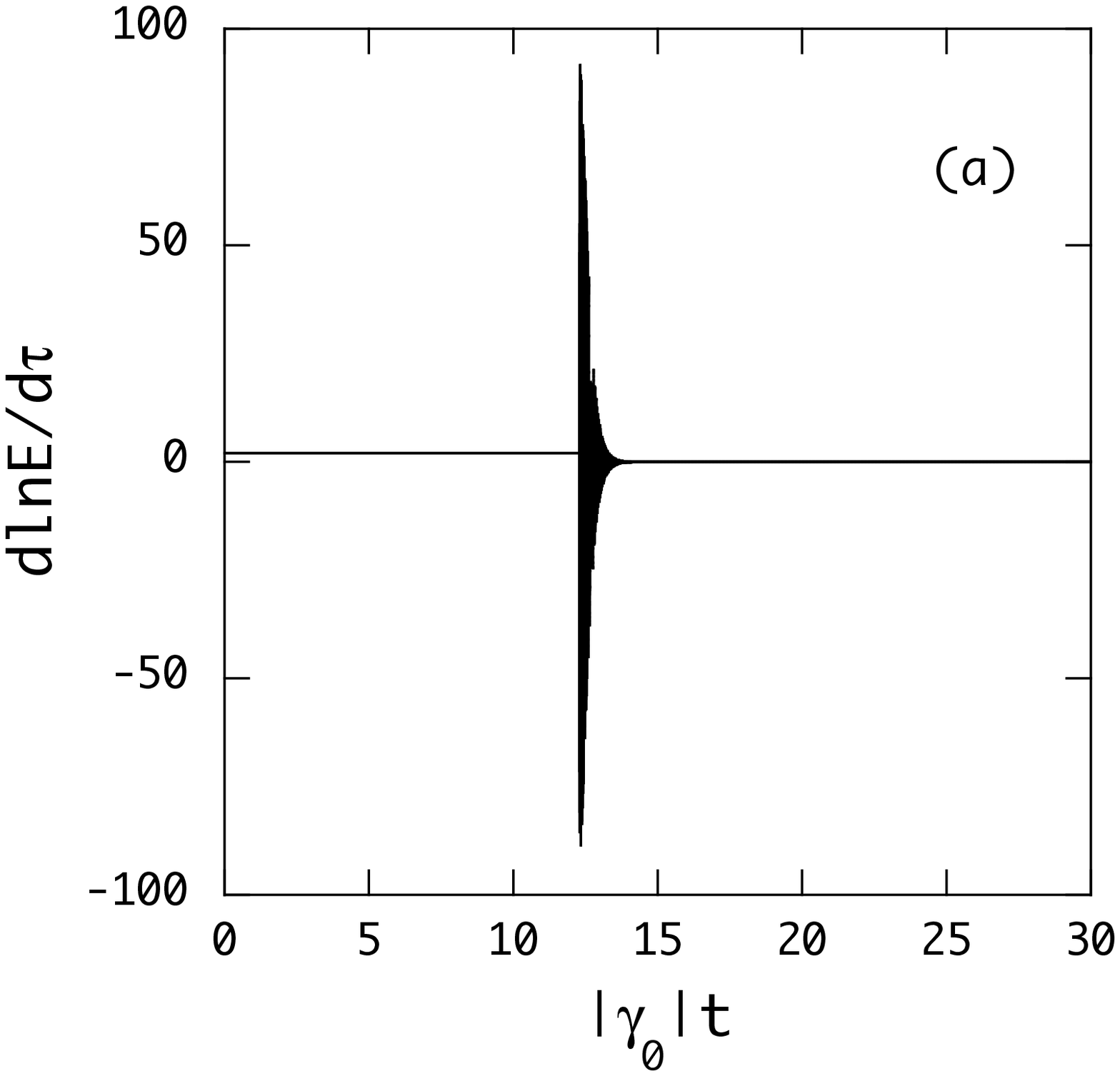}}
\resizebox{0.5\columnwidth}{!}{
\includegraphics{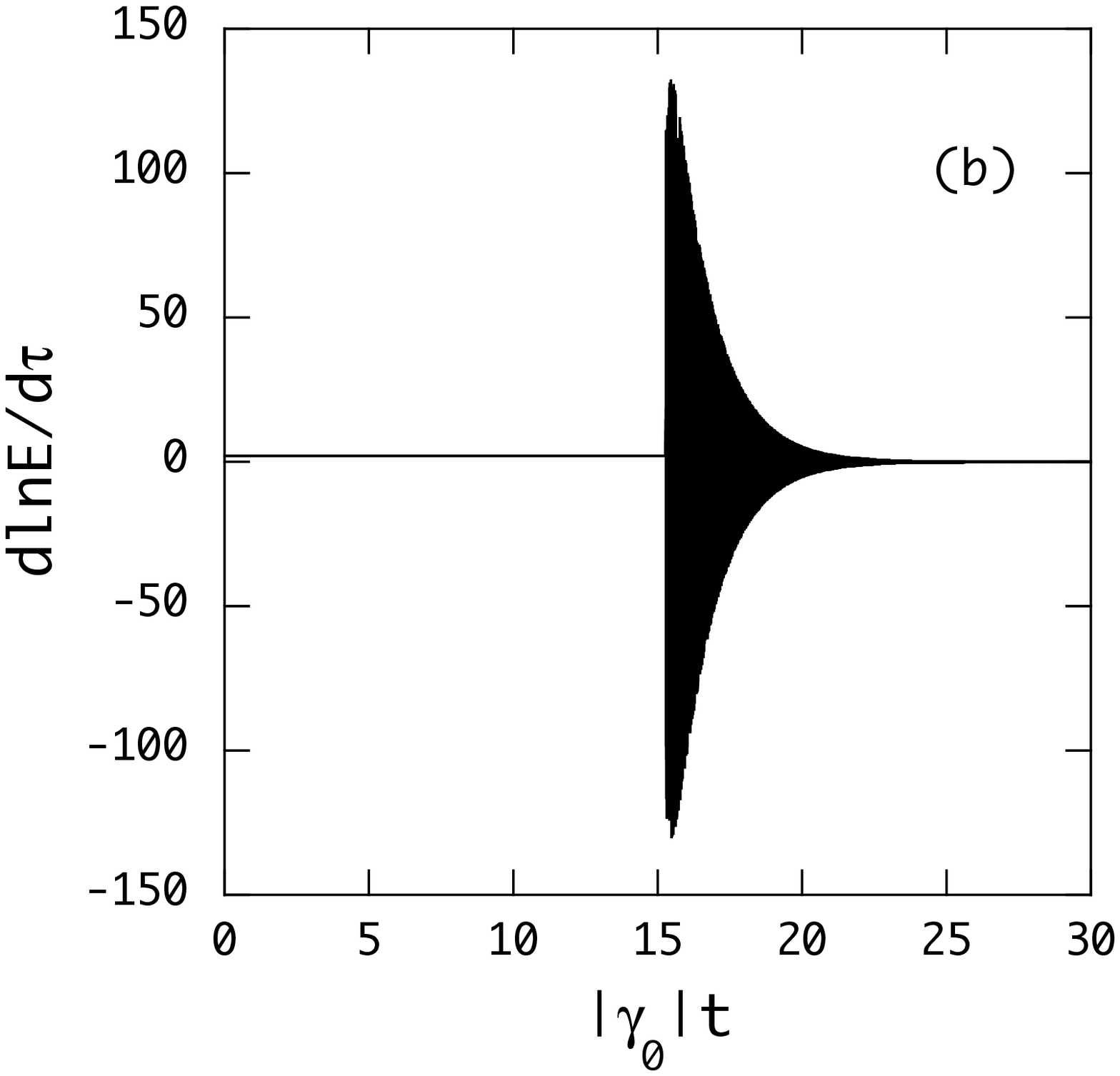}}
\resizebox{0.5\columnwidth}{!}{
\includegraphics{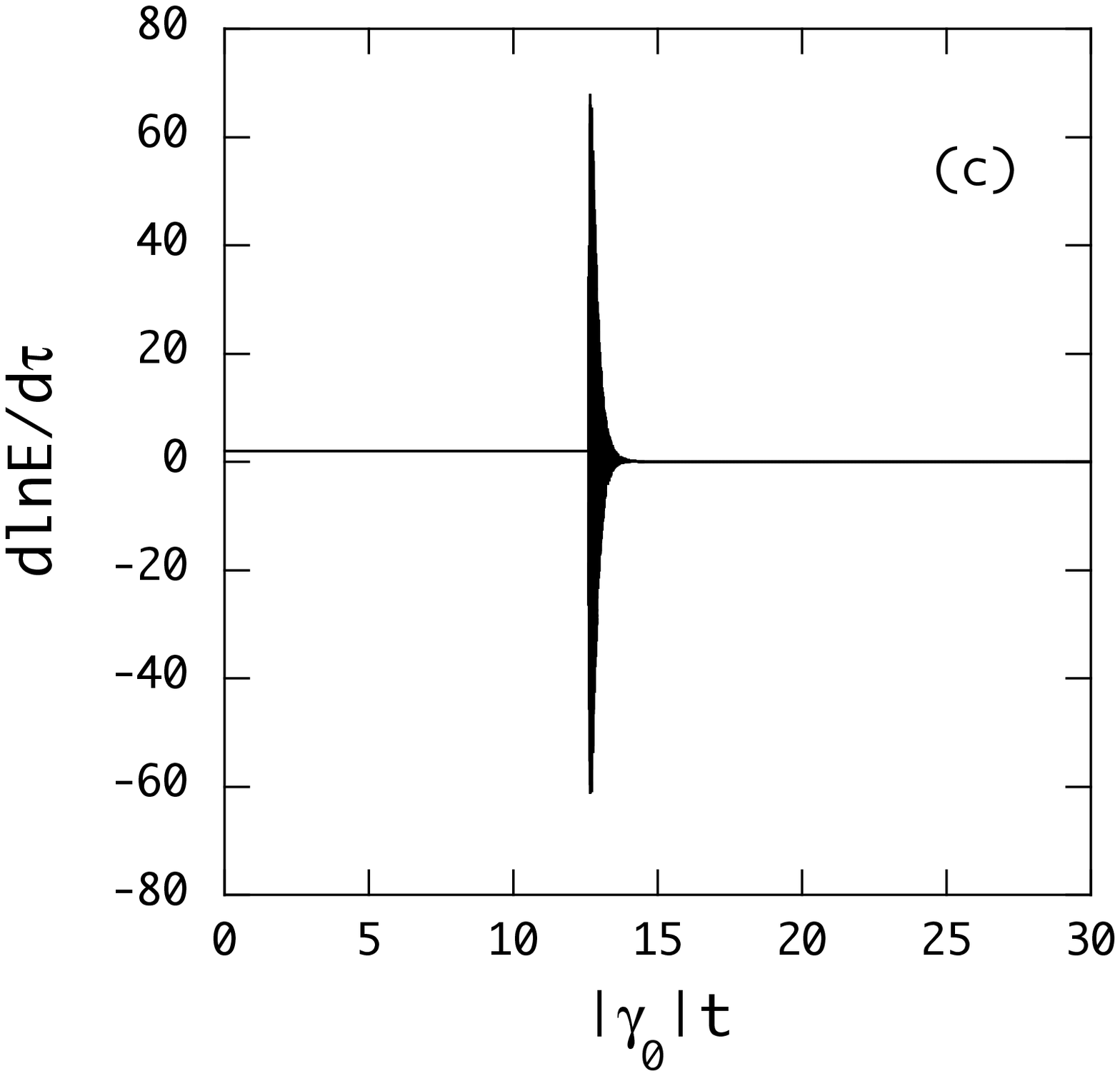}}
\resizebox{0.5\columnwidth}{!}{
\includegraphics{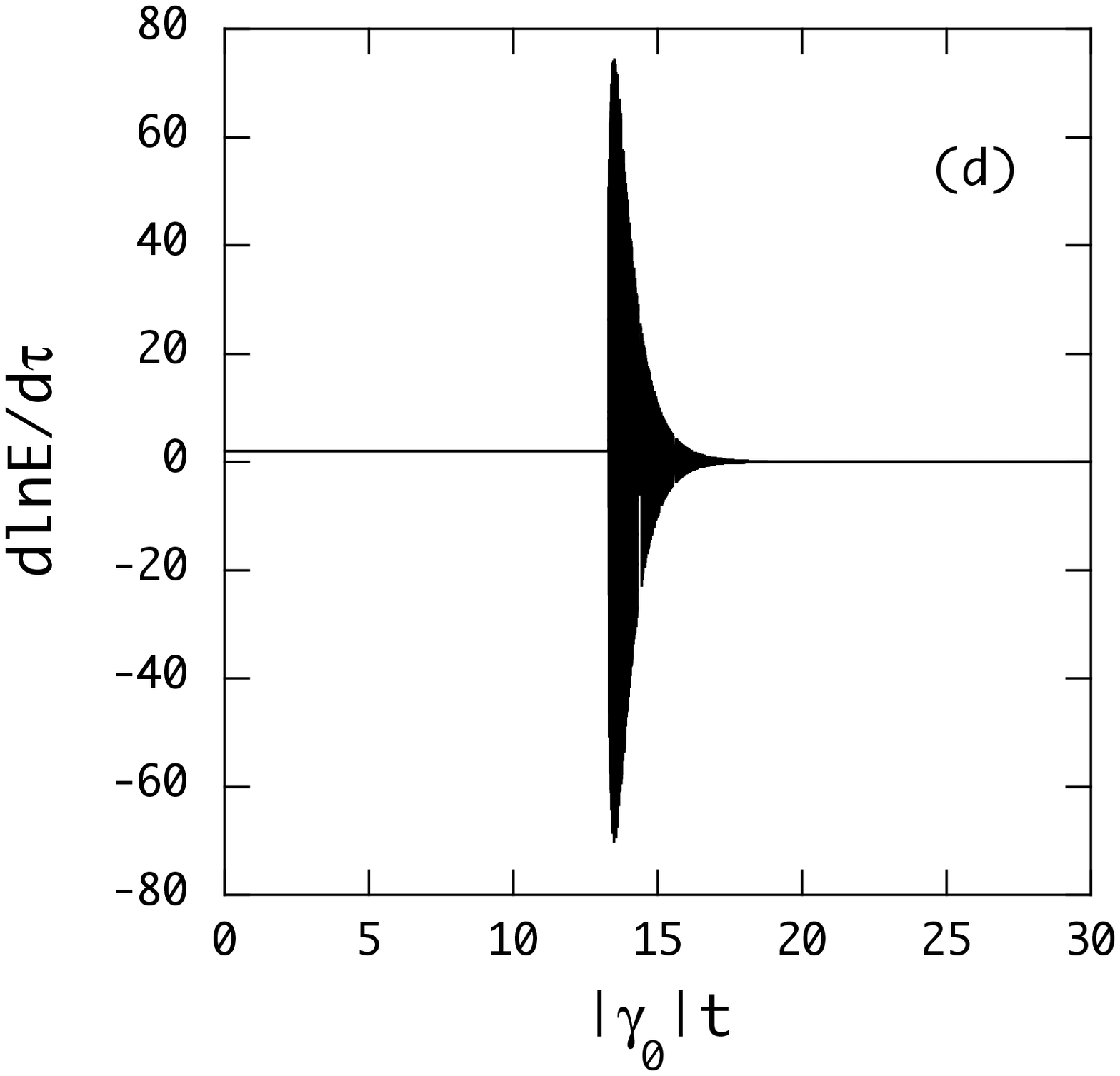}}
\caption{Same as Fig. \ref{fig:amp_thmc} but for $d\ln E/d\tau$ versus $\tau=|\gamma_0|t$ where $E=|a|^2+|b|^2+|c|^2$.
}
\label{fig:DE_thmc}
\end{figure*}

\begin{figure}
\resizebox{0.49\columnwidth}{!}{
\includegraphics{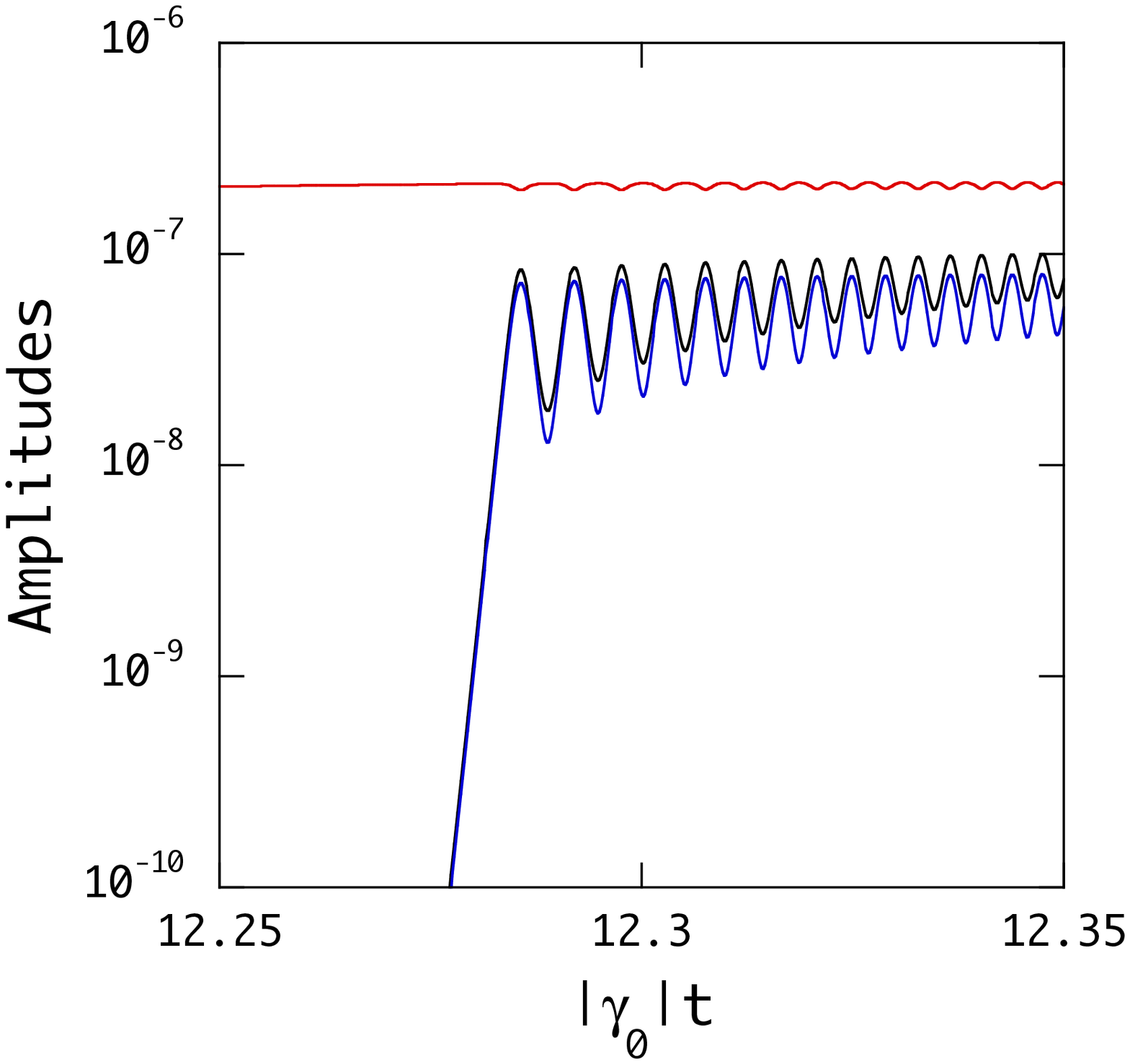}}
\resizebox{0.49\columnwidth}{!}{
\includegraphics{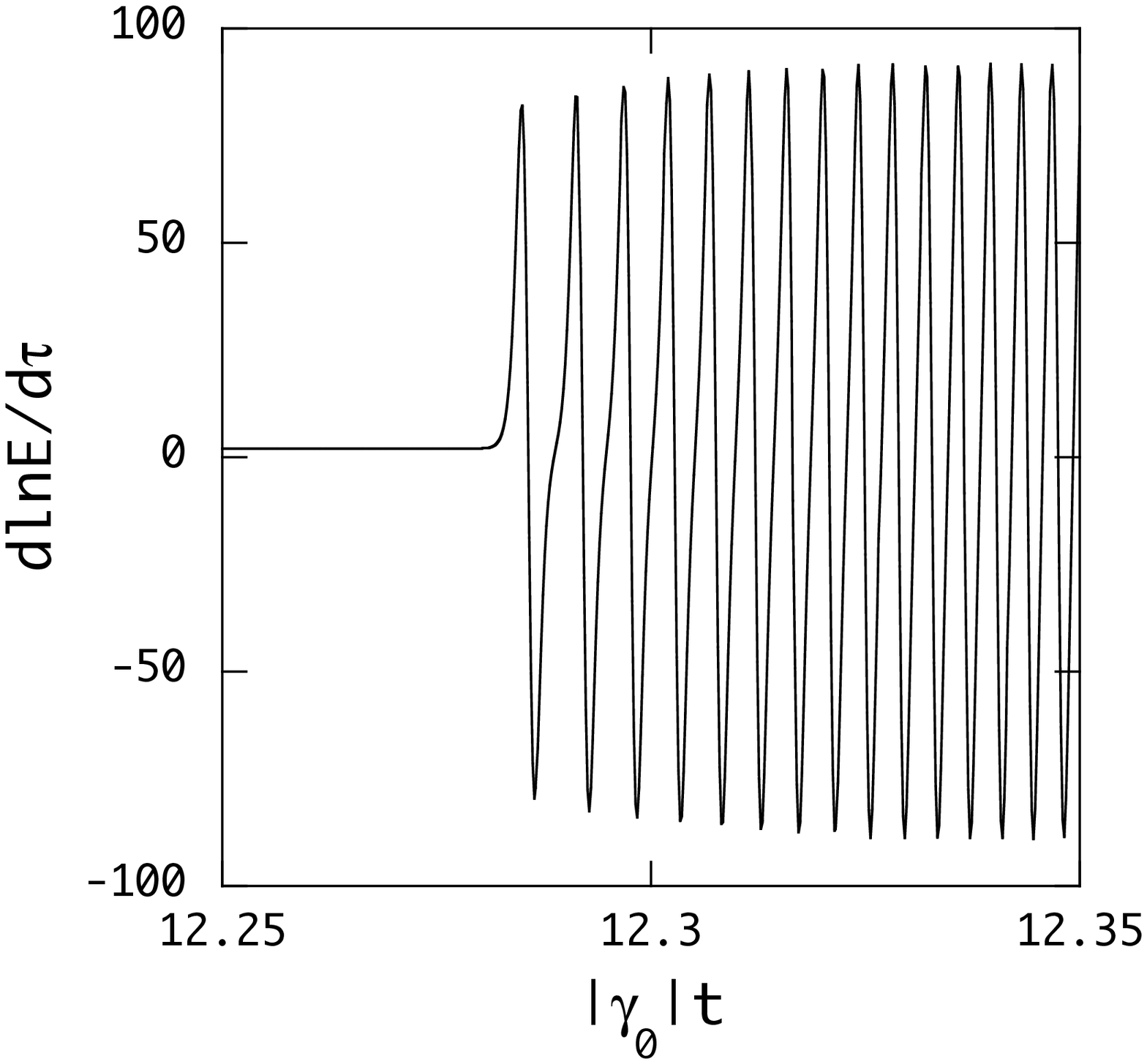}}
\caption{Mode amplitudes and $d\ln E/d\tau$ versus $\tau=|\gamma_0|t$ at the phase of the
daughter mode excitation
where the OsC mode $B_1$ is coupled with $(g_{60}^{-1},g_{60}^{+2})$ modes.
}
\label{fig:enlarge}
\end{figure}

\begin{figure}
\resizebox{0.49\columnwidth}{!}{
\includegraphics{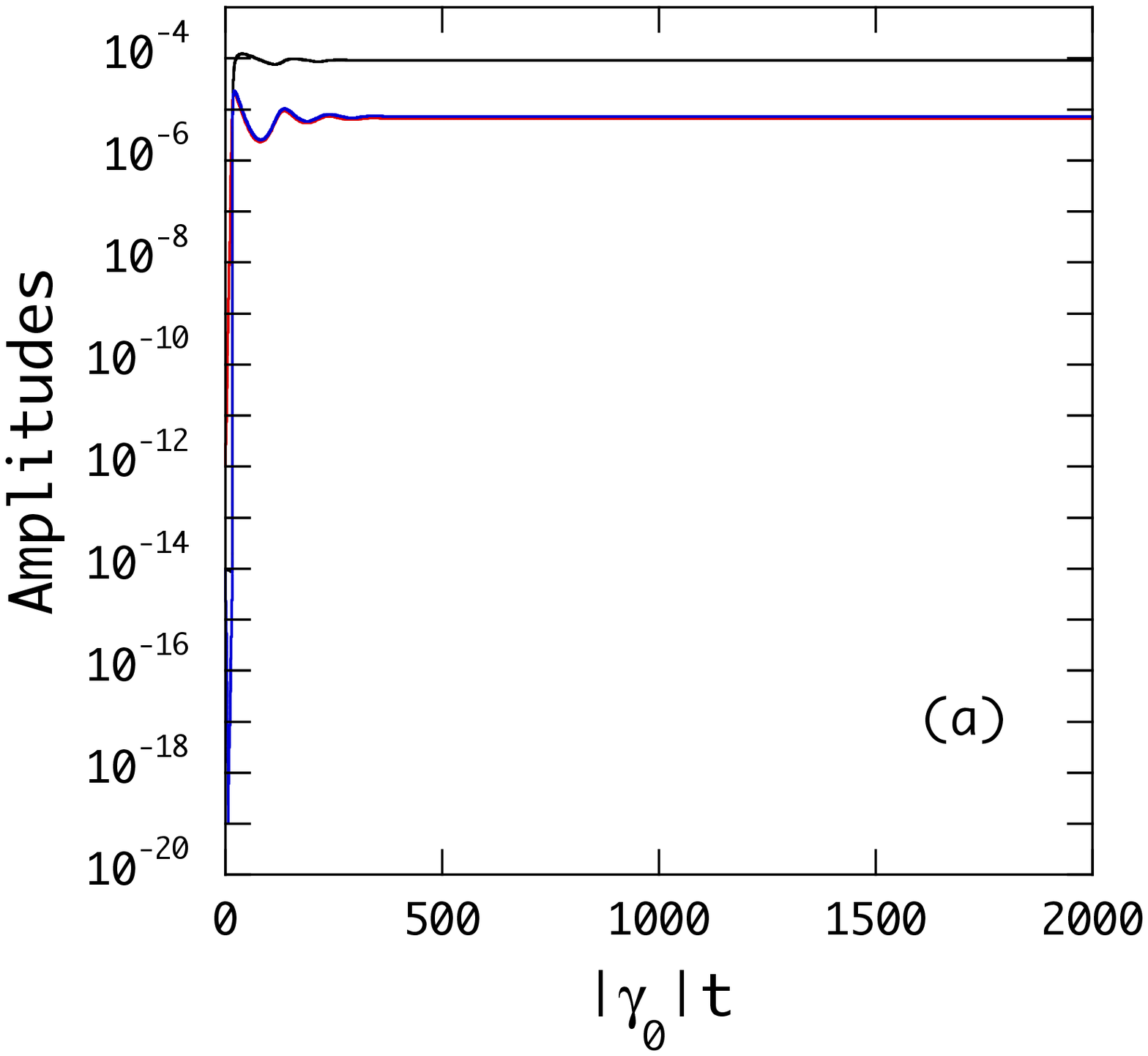}}
\resizebox{0.49\columnwidth}{!}{
\includegraphics{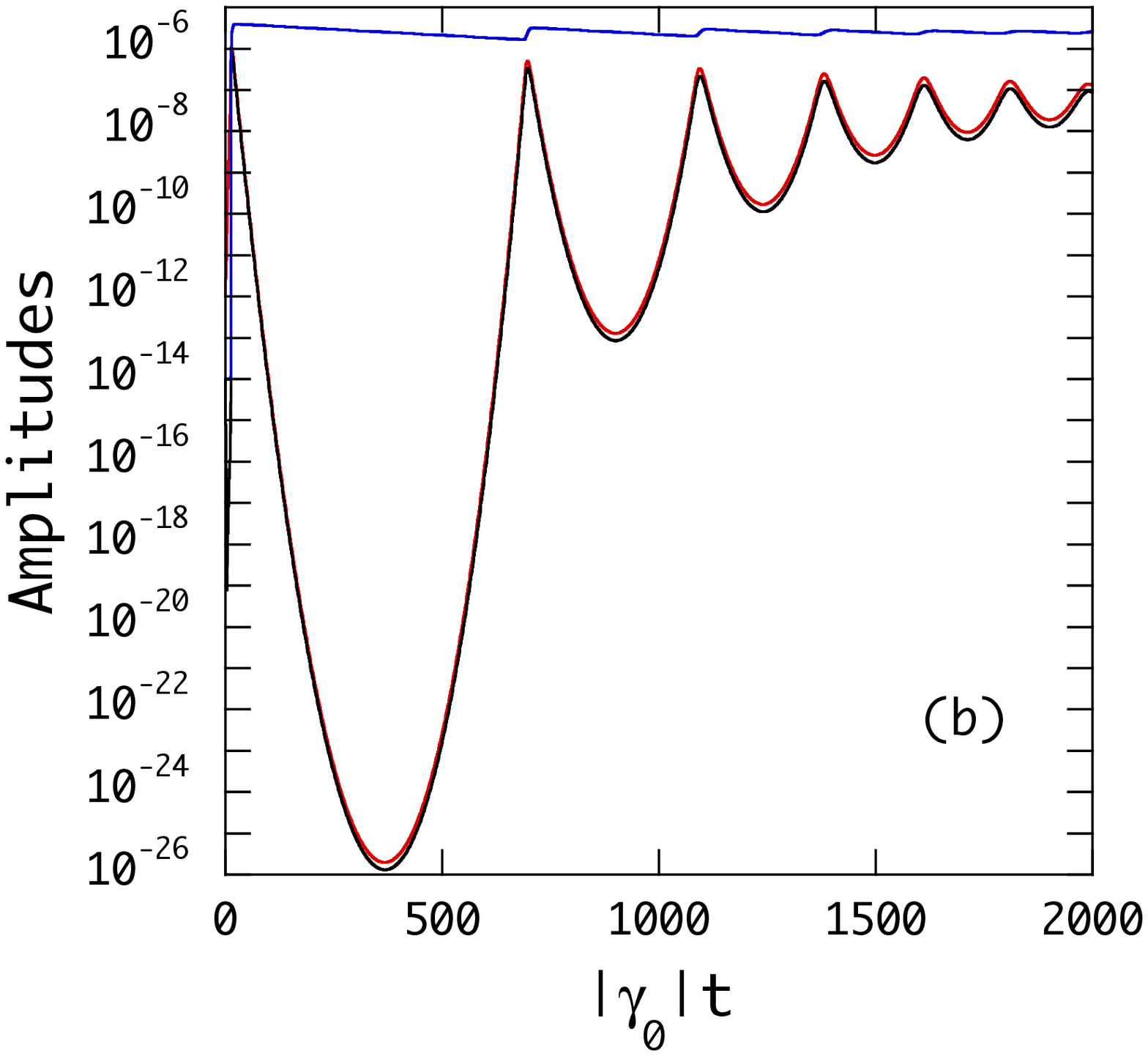}}
\caption{Mode amplitudes versus $\tau=|\gamma_0|t$ shown for a long span of $\tau$ for the case where the OsC mode $B_1$ is coupled with $(g_{15}^{-1},g_{60}^{+2})$ modes (panel a) 
and $(g_{60}^{-1},r_{15}^{+2})$ modes (panel b).
}
\label{fig:longspan}
\end{figure}

\begin{figure}
\resizebox{0.49\columnwidth}{!}{
\includegraphics{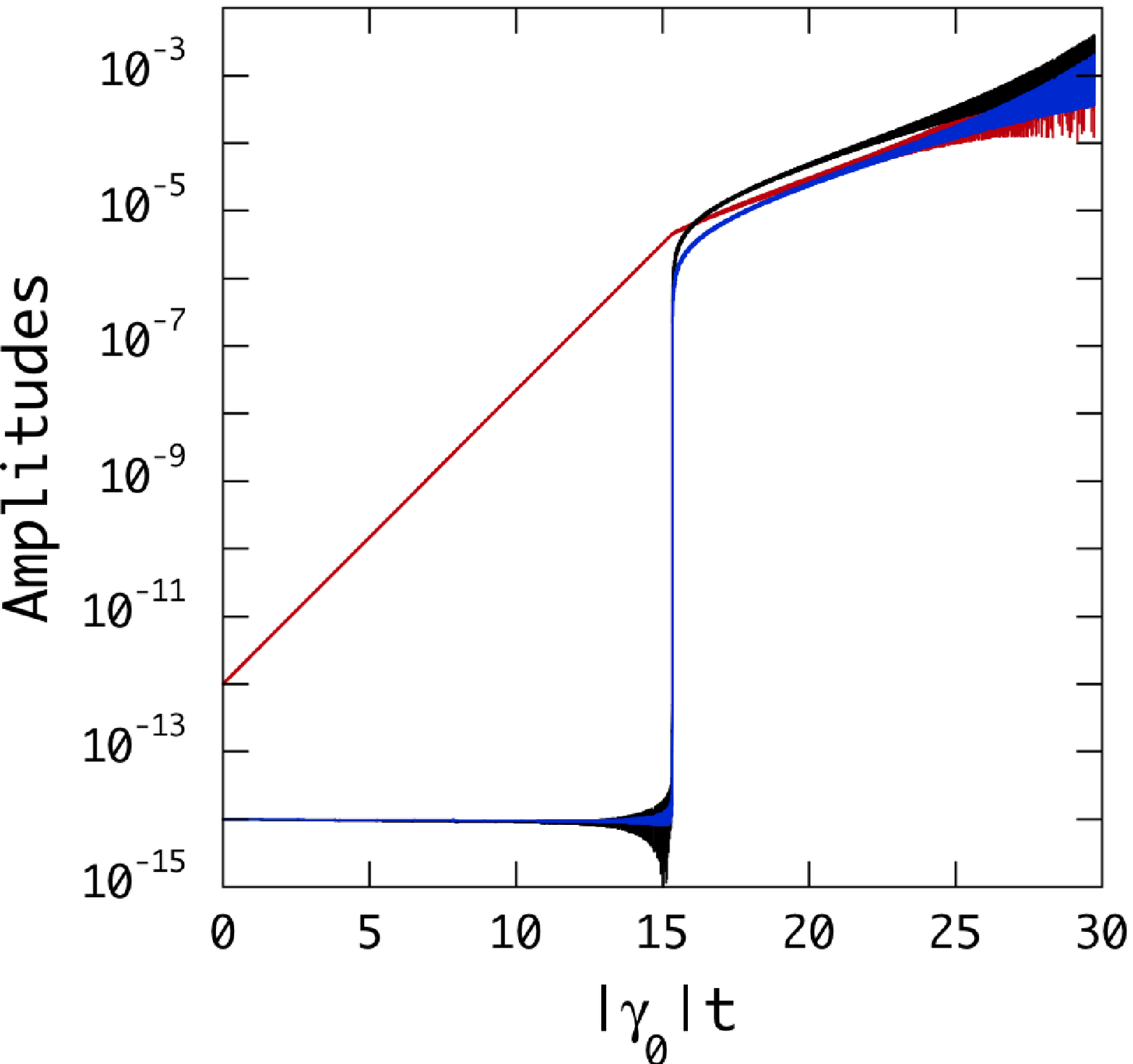}}
\resizebox{0.49\columnwidth}{!}{
\includegraphics{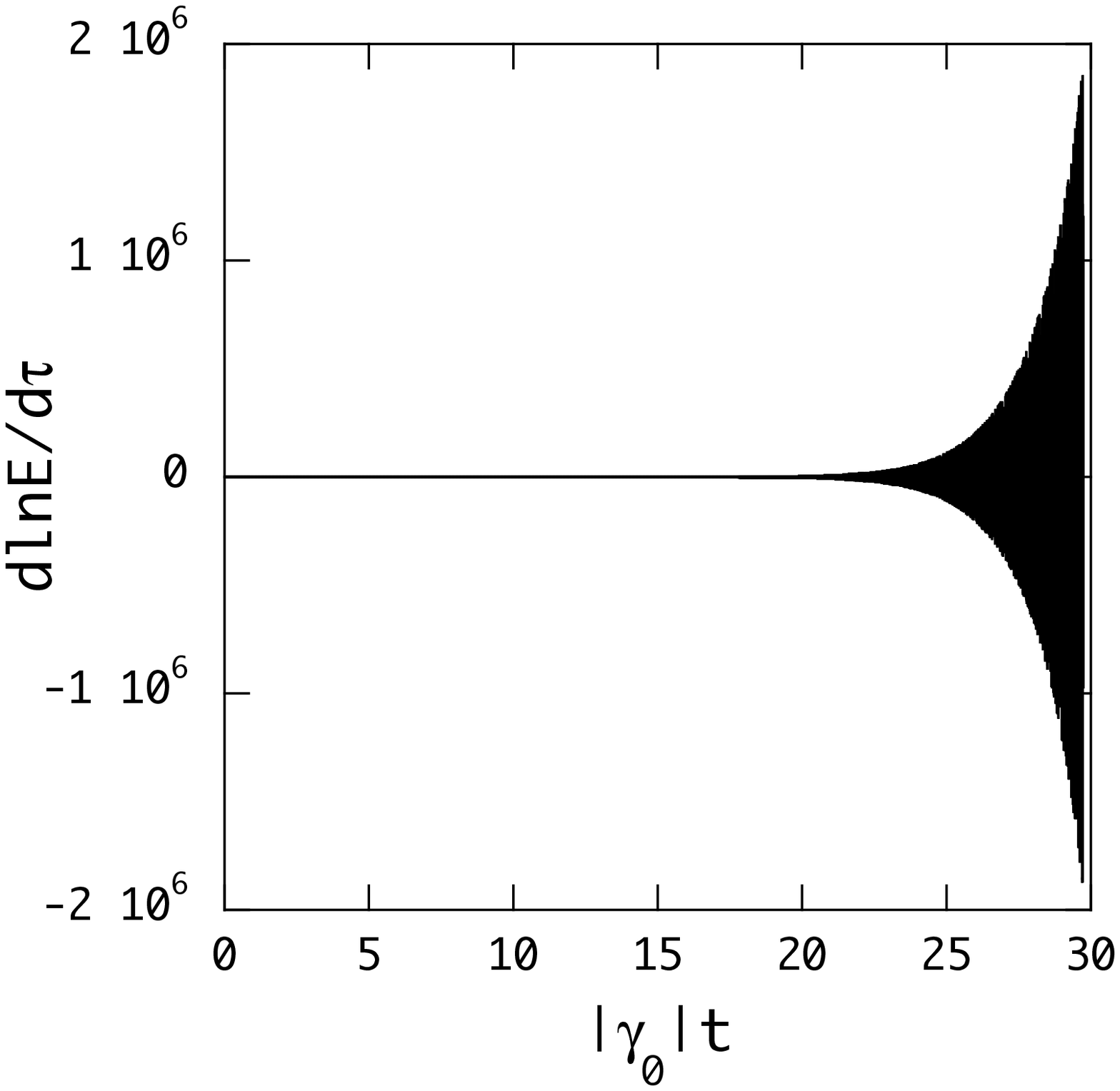}}
\caption{Mode amplitudes and $d\ln E/d\tau$ versus $\tau=|\gamma_0|t$ for 
the OsC mode $B_1$ coupled to $(g_{15}^{-1},g_{15}^{+2})$.
}
\label{fig:g15r15}
\end{figure}

If the parent mode is an OsC mode of $m_{\rm OsC}<0$, 
we have ${\rm Re}(\sigma_{\rm OsC})>0$, 
$s_{\rm OsC}<0$, and $\gamma_{\rm OsC}<0$ and hence $Q_a=Q_{\rm OsC}>0$.
Therefore, for the daughter modes $b$ and $c$, 
the conditions $Q_b>0$ and $Q_c>0$ implies that $s_b{\rm Re}(\sigma_b)>0$ and $s_c{\rm Re}(\sigma_c)>0$ have to be satisfied to obtain a stable equilibrium state of amplitudes.
On the other hand, to satisfy the selection rule (\ref{eq:select_m}) for $m_a=m_{\rm OsC}<0$,
at least one of $m_b$ and $m_c$ must be positive.
If $m_b<0$ and $m_c>0$, for example, the daughter modes with $m_b$ should be prograde modes and those with $m_c$ retrograde modes, for which we may consider both $g$-modes and $r$-modes.

The properties of the amplitude evolutions caused by non-linear three-mode couplings have been extensively investigated
(e.g., \citealt{WersingerFinnOtt1980}).
However, it is of some interest to show a few examples of the evolutions obtained for the low frequency oscillation modes of rotating stars.
Figs. \ref{fig:amp_thmc} and \ref{fig:DE_thmc}
show some examples of amplitude evolutions of even $g$- and $r$-modes
driven by the $m=-1$ OsC mode $B_1$ through non-linear three-mode
coupling where the initial amplitudes of the modes for numerical integration are given by $a=10^{-12}$ for the OsC mode and by $b=c=10^{-14}$ for the daughter modes.
Although these initial amplitudes are assumed somewhat arbitrarily,
we confirm that the final states of the amplitude evolutions do not strongly depend on
the initial conditions so long as the initial amplitudes of the parent and
daughter modes are much smaller than those expected for equilibrium states and the daughter modes
have much smaller initial amplitudes than the parent mode.
For example, we have tried the initial amplitudes $a=10^{-10}$ and $b=c=10^{-12}$
and found that for $a=10^{-10}$ the parent modes reach the critical amplitude of non-linear excitation of the daughter modes earlier than for $a=10^{-12}$ but
the final states of the amplitude evolutions are almost the same as those for
the initial amplitudes $a=10^{-12}$ and $b=c=10^{-14}$.
Table \ref{tab:modedata} tabulates the frequency of the OsC mode, $g$- and $r$-modes and their growth rate and damping rates used for the amplitude evolution calculations.
Table \ref{tab:gammaratio}, on the other hand, tabulates $\Delta\overline{\gamma}_{abc}=\Delta\gamma_{abc}/\sigma_0$, 
$(\gamma_b+\gamma_c)/|\gamma_a|$, and $\gamma_b/\gamma_c$, which may be useful to understand the numerical results 
discussed below.

\begin{table}
\begin{center}
\caption{Frequency $\overline\sigma_{\rm R}$ and the growth rate $\overline\gamma$ of 
the low frequency modes in the three-mode couplings shown in the figures \ref{fig:amp_thmc} and \ref{fig:g15r15}.}
\begin{tabular}{ccrrc}
\hline
$m$  & mode & $\overline\sigma_{\rm R}$ & $\overline\gamma$ & $s$ \\
\hline
  $-1$ & $B_1$ & 0.2206 & $-6.21(-5)$ & $-1$\\
 \hline
  $-1$ & $g_{15}$ & 0.3743 & $5.68(-7)$ & +1\\
  $-1$ & $g_{60}$ & 0.2420 & $1.54(-4)$ & +1\\
 \hline
  +2 & $g_{15}$ & 0.07435 & $4.74(-7)$ & +1\\
  +2 & $g_{60}$ & $-0.1842$ & $3.83(-4)$ & $-1$\\
  +2 & $r_{15}$ & $-0.3443$ & $8.19(-8)$ & $-1$\\
  +2 & $r_{60}$ & $-0.3764$ & $8.97(-5)$ & $-1$\\
\hline
\end{tabular}
\label{tab:modedata}
\end{center}
\end{table}

\begin{table}
\begin{center}
\caption{$\Delta\overline\gamma_{abc}=\Delta\gamma_{abc}/\sigma_0$ and the ratios $(\gamma_b+\gamma_c)/|\gamma_a|$ and $\gamma_b/\gamma_c$
for three-mode coupling for the $2M_\odot$ ZAMS model at $\overline\Omega_s=0.2$ where
the modes in the left most column are 
coupled to OsC $B_1$ mode of $m_a=-1$.}
\begin{tabular}{cccc}
\hline
daughter modes  & $\Delta\overline{\gamma}_{abc}$ & $(\gamma_b+\gamma_c)/|\gamma_a|$ & $\gamma_b/\gamma_c$ \\
\hline
($g_{60}^{-1}$,$g_{60}^{+2}$) &4.85($-$4) & 10.2 & 0.402\\
($g_{60}^{-1}$,$r_{60}^{+2}$) &1.91($-$4) & 4.64 & 1.71\\
($g_{15}^{-1}$,$g_{60}^{+2}$) &3.31($-$4) & 7.31 &1.48($-$3)\\
($g_{60}^{-1}$,$r_{15}^{+2}$) &1.02($-$4) & 2.93 & 1.88(+3)\\
($g_{15}^{-1}$,$g_{15}^{+2}$) &$-$6.11($-$5) & 0.0168 & 1.20\\
\hline
\end{tabular}
\label{tab:gammaratio}
\end{center}
\end{table}

Fig. \ref{fig:amp_thmc} shows that
the parent mode grows and the daughter modes decay almost exponentially
with increasing $\tau=|\gamma_0|t$ from $\tau=0$ where
$\gamma_0=\gamma_a$.
The panels (a) and (b) correspond to the cases where the daughter modes have damping rates
comparable to or larger than the growth rate $|\gamma_0|$ while
the panels (c) and (d) show the cases where one of the daughter modes has a very small damping rate
compared to the other, which has the damping rate comparable to or greater than $|\gamma_0|$.
The small damping rate mode
describes almost a horizontal line in the early periods of evolution from $\tau=0$.
Because of the initial conditions, the OsC mode is dominant over the daughter modes when $\tau\sim0$ and hence
$d\ln E/d\tau\approx 2$ with $E=|a|^2+|b|^2+|c|^2$ as shown by Fig. \ref{fig:DE_thmc}.

When the amplitude of the parent mode exceeds some threshold value, the amplitudes of
the daughter modes start rising abruptly to become comparable to the parent mode.
This behavior may be understood by rewriting, for example, equation (\ref{eq:dothatb}) as
\be
{\dot {\hat b}/ \hat b}=-\gamma_b-2{\rm Re}(\sigma_b)s_b|\eta_{abc}|\hat a({\hat c/ \hat b})\sin\varphi,
\label{eq:dotbb}
\ee
which suggests that the evolution of $\hat b$ can be strongly affected when
$2{\rm Re}(\sigma_b)s_b|\eta_{abc}|\hat a\gtrsim1$ and $\hat c/\hat b\sim 1$ even if $\hat b$ and $\hat c$
are much smaller than $\hat a$ (see Fig. \ref{fig:amp_thmc}).
Equation (\ref{eq:dotbb}) indicates that if $-2{\rm Re}(\sigma_b)s_b|\eta_{abc}|\hat a(\hat c/\hat b)\sin\varphi>\gamma_b$ is satisfied
for $\hat a$ greater than the critical value $\hat a_{\rm crit}$,
the amplitude $\hat b$ grows rapidly.
Note that since ${\rm Re}(\sigma_b)s_b>0$ for $g$-modes and $r$-modes (see Table 3), 
we need $\sin\varphi<-\gamma_b/(2{\rm Re}(\sigma_b)s_b|\eta_{abc}|\hat a\hat c/\hat b)<0$ for the rapid growth of the amplitudes $\hat b$ and $\hat c$.
Fig.\ref{fig:enlarge} plots the amplitudes $a$, $b$, and $c$ and the derivative $d\ln E/d\tau$
as a function of $\tau$ in the period of abrupt amplitude rise of the daughter modes.

Fig. \ref{fig:DE_thmc} also suggests that immediately after the rapid amplitude rise,
$d\ln E/d\tau$ fluctuates with short periods and the fluctuation amplitudes quickly damp, 
suggesting that the mode amplitudes of the parent and daughter modes stay finite.
The rapid fluctuations of $d\ln E/d\tau$ is caused by the small amplitude fluctuations of the mode amplitudes.

\begin{figure*}
\resizebox{0.66\columnwidth}{!}{
\includegraphics{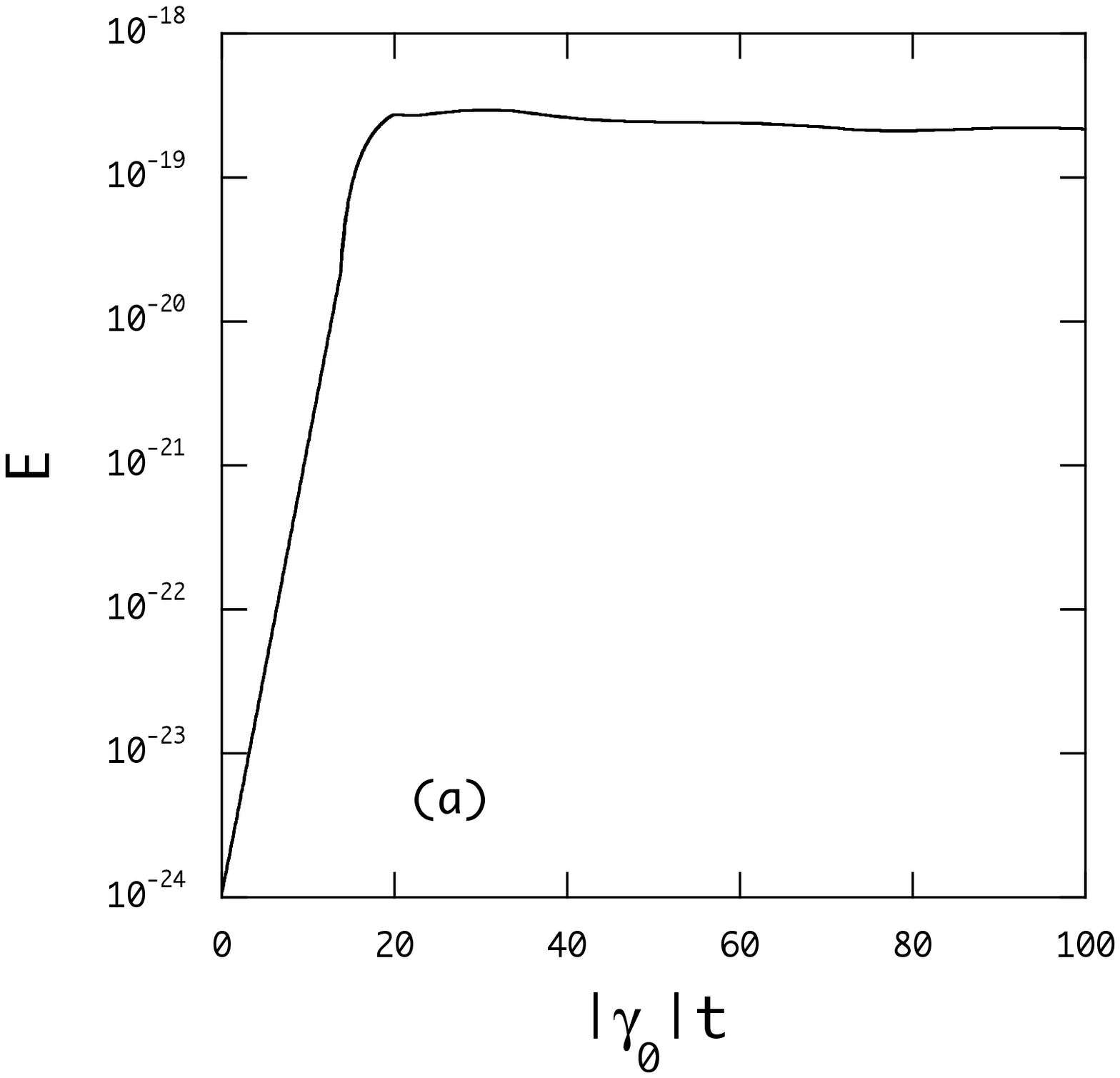}}
\resizebox{0.66\columnwidth}{!}{
\includegraphics{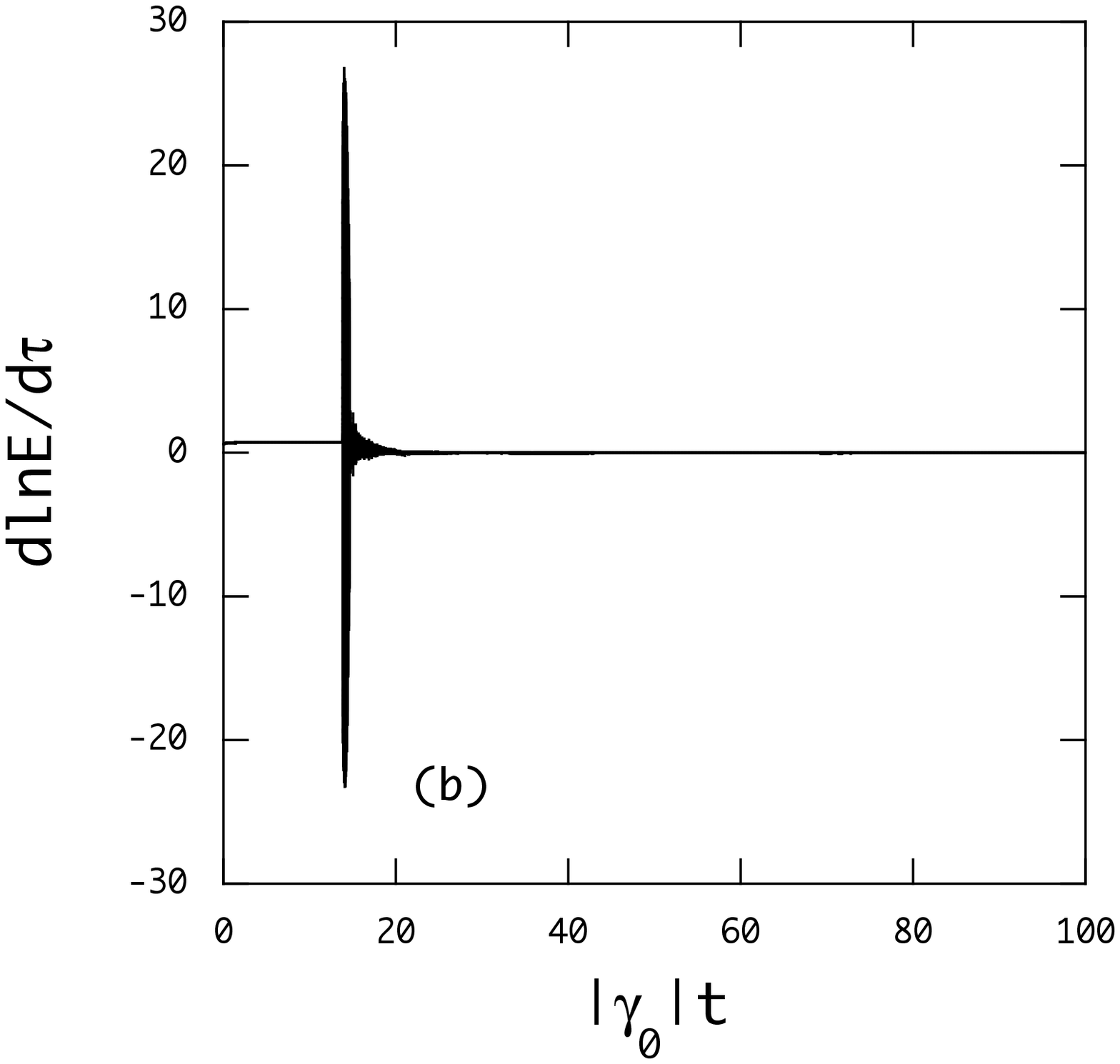}}
\resizebox{0.66\columnwidth}{!}{
\includegraphics{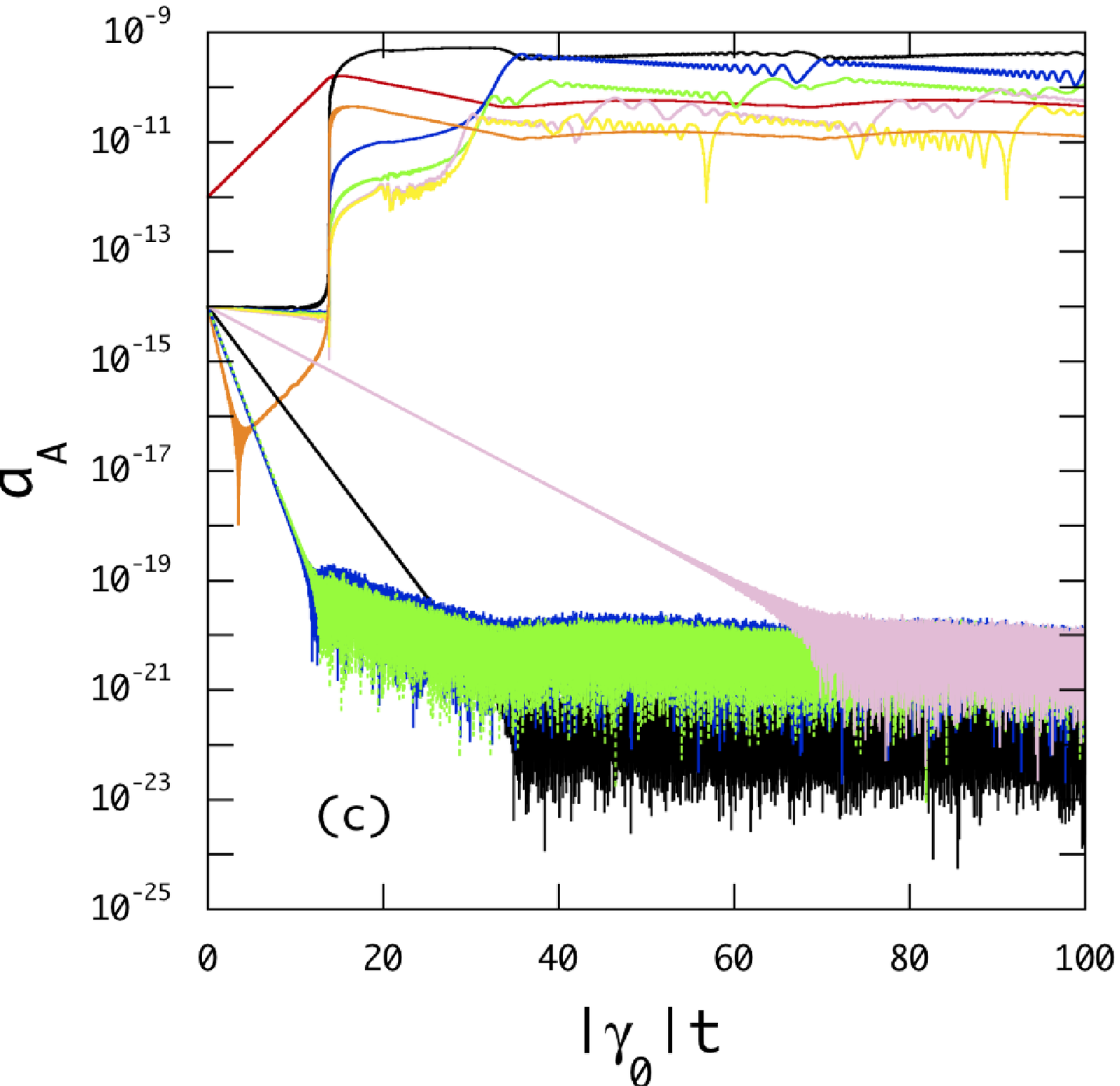}}
\caption{$E=|a_1|^2+\sum_j|b_j|^2+\sum_k|c_k|^2$ (panel a), $d\ln E/d\tau$ (panel b), and $a_A$ of 
smallest and largest magnitudes (panel c) plotted as a function of $\tau=|\gamma_0|t$ where $|\overline\gamma_0|=1.71\times10^{-4}$. In the panel (c), the red line stands for the $m_a=-1$ OsC mode $B_1$ and
other lines of different colors for the small and large amplitude modes.
}
\label{fig:a_mm1_B1}
\end{figure*}

\begin{figure}
\resizebox{1.\columnwidth}{!}{
\includegraphics{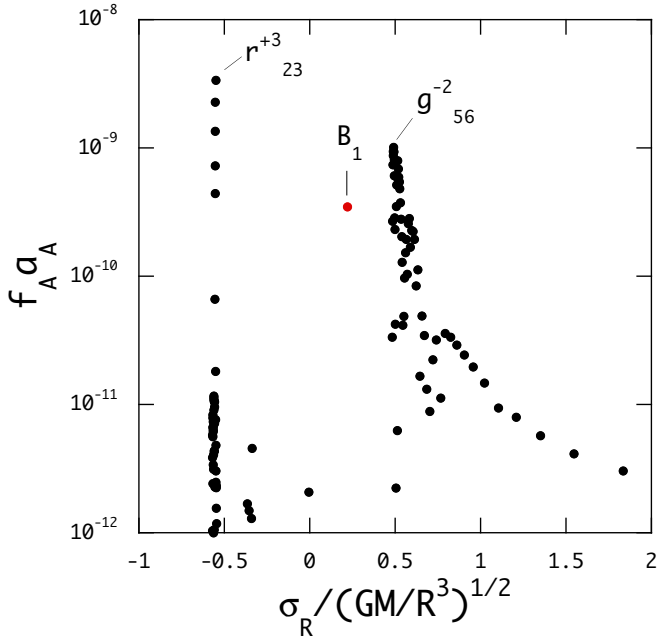}}
\caption{Amplitudes $f_Aa_A$ of non-linearly excited low frequency modes versus the frequency $\overline\sigma_{\rm R}$ for the case in which
the OsC mode $B_1$ of $m_a=-1$ is the parent mode where the filled red circle indicates the OsC mode $B_1$ of $m_a=-1$.
}
\label{fig:faaa_B1_mm1}
\end{figure}

Evolution of the mode amplitudes after the three modes have comparable amplitudes depends 
on the three quantities $\Delta\gamma_{abc}$, $(\gamma_b+\gamma_c)/\gamma_a$, and $\gamma_b/\gamma_c$.
When $\Delta\gamma_{abc}>0$, $(\gamma_b+\gamma_c)/|\gamma_a|\sim1$, and $\gamma_b/\gamma_b\sim1$,
the amplitudes $a$, $b$, and $c$ quickly converge to equilibrium amplitudes.
On the other hand, when $\Delta\gamma_{abc}>0$, $(\gamma_b+\gamma_c)/|\gamma_a|\sim1$, but $\gamma_b/\gamma_c\gg 1$ (or $\gamma_b/\gamma_c\ll 1$),
evolution of the amplitudes shows much more complicated behavior, as shown by Fig. \ref{fig:longspan}.
If the OsC mode is coupled to $(g_{15}^{-1},g_{60}^{+2})$ (panel a), 
the smaller damping rate mode
$g_{15}^{-1}$ soon attains a constant amplitude, but it takes 
much longer time for the other two modes to reach equilibrium amplitudes.
For the coupling to $(g_{60}^{-1},r_{15}^{+2})$ (panel b), on the other hand, although the mode amplitudes
are kept finite, they fluctuate with long periods and
their amplitudes of fluctuations only gradually decrease as $\tau\rightarrow\infty$.

There are cases in which the mode amplitudes diverge as $\tau$ increases.
An example of such amplitude evolution is given by Fig. \ref{fig:g15r15} where the OsC mode is coupled to 
two small damping rate modes $g_{15}^{-1}$ and $r_{15}^{+2}$.
For this coupling, we have $\Delta\gamma_{abc}<0$, $(\gamma_b+\gamma_c)/|\gamma_a|\ll1$, and $\gamma_b/\gamma_b\sim1$.
After the amplitudes $a$, $b$, and $c$ become comparable, the amplitudes and the derivative $d\ln E/d\tau$
start to fluctuate with very short periods and their fluctuation amplitudes steadily increase with $\tau$, suggesting
that the evolution may not converge to states of finite amplitudes when  $\tau\rightarrow\infty$.

\begin{figure*}
\resizebox{0.66\columnwidth}{!}{
\includegraphics{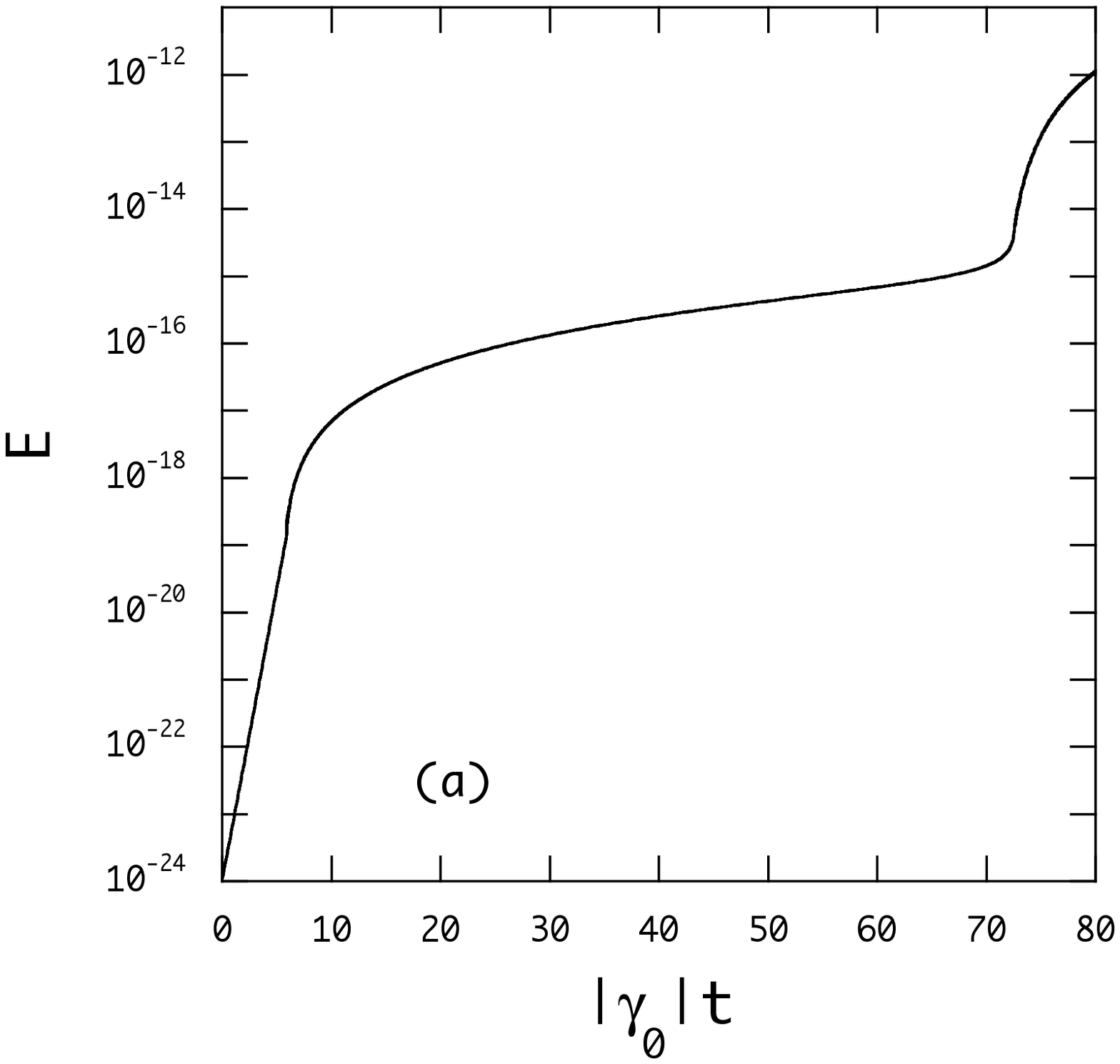}}
\resizebox{0.66\columnwidth}{!}{
\includegraphics{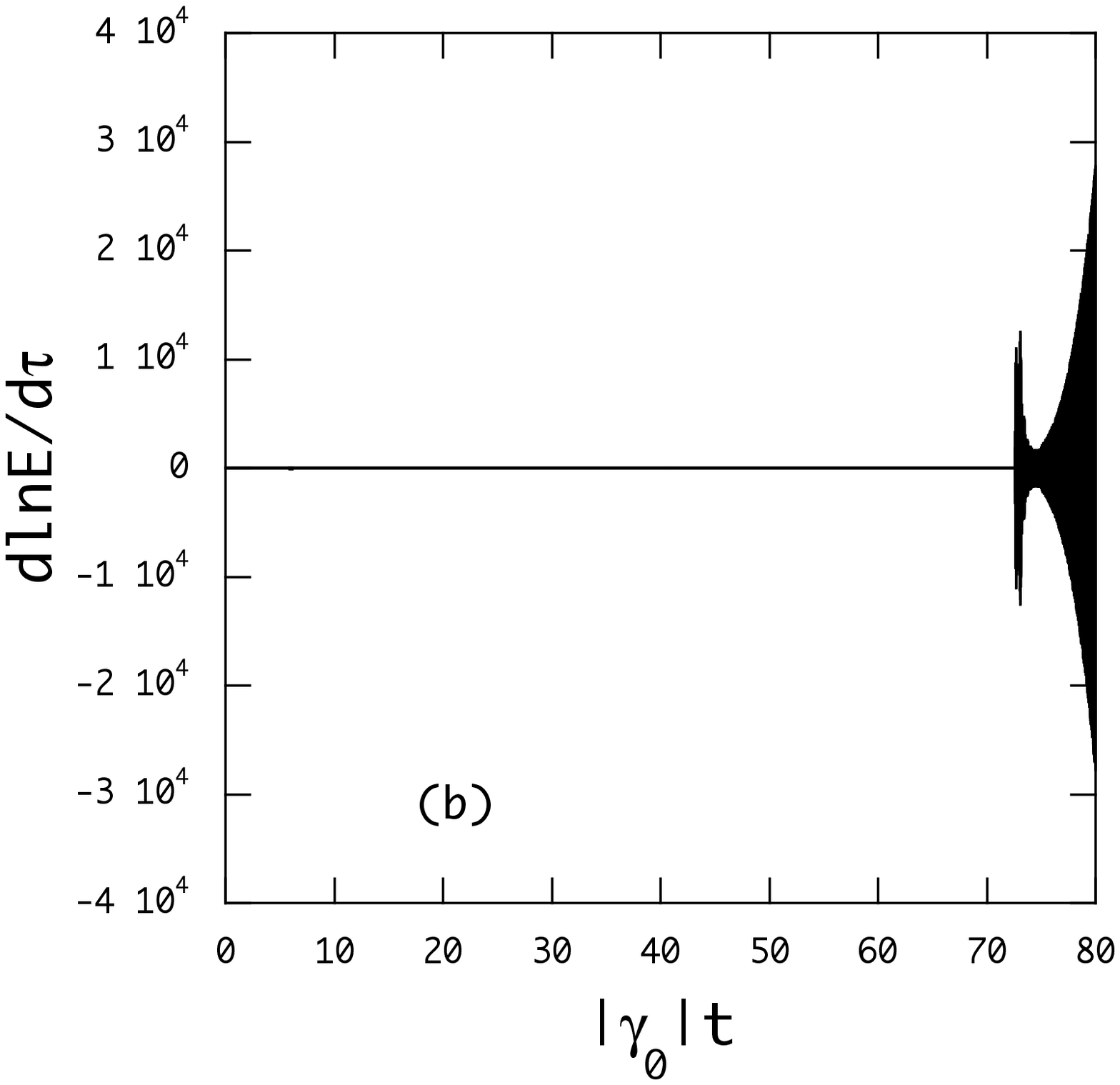}}
\resizebox{0.66\columnwidth}{!}{
\includegraphics{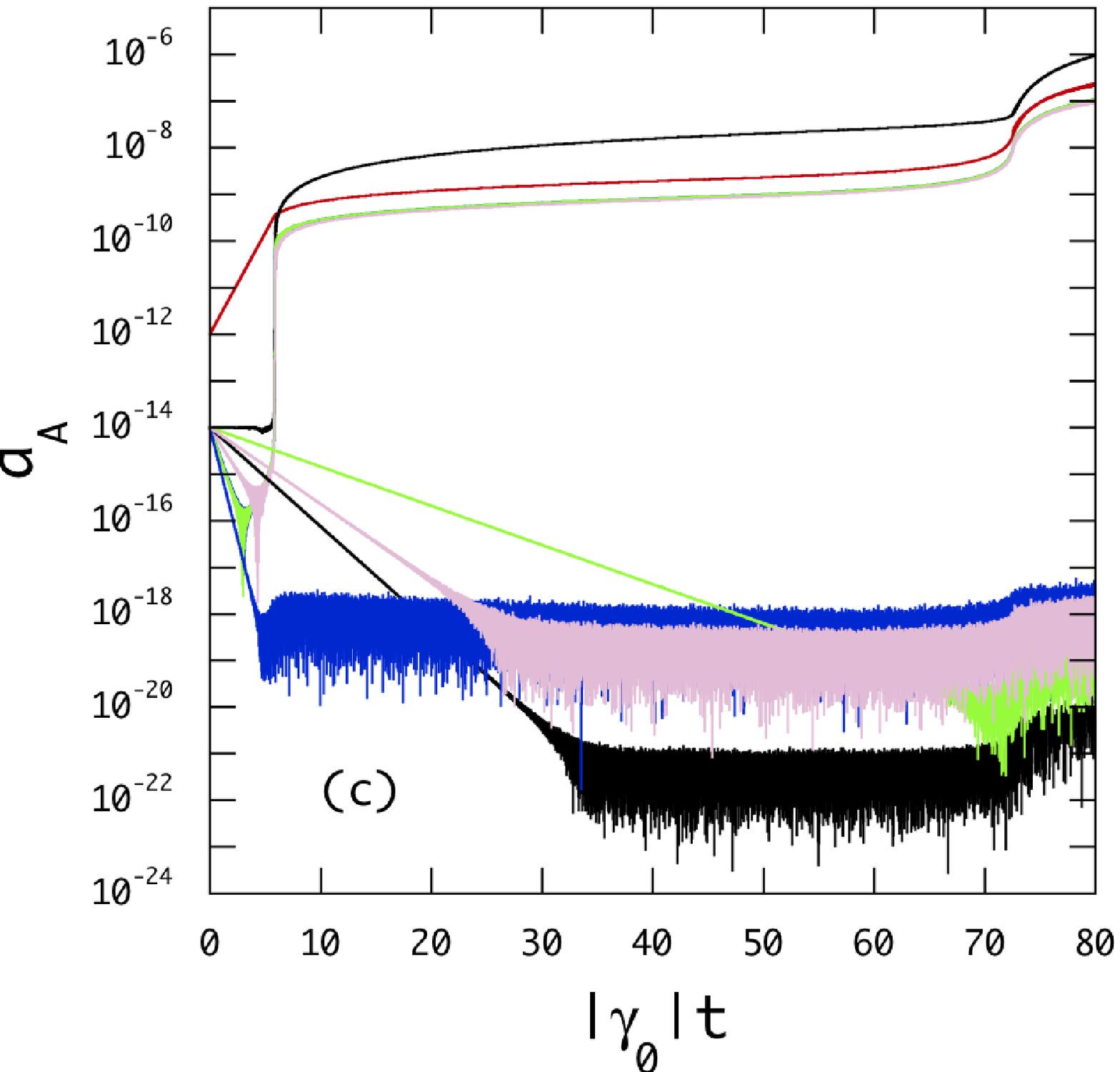}}
\caption{Same as Fig.\ref{fig:a_mm1_B1} but for the case in which the OsC mode $B_2$ is the parent mode
where $|\overline\gamma_0|=1.71\times10^{-4}$.
}
\label{fig:a_mm1_B2}
\end{figure*}

\begin{figure*}
\resizebox{0.66\columnwidth}{!}{
\includegraphics{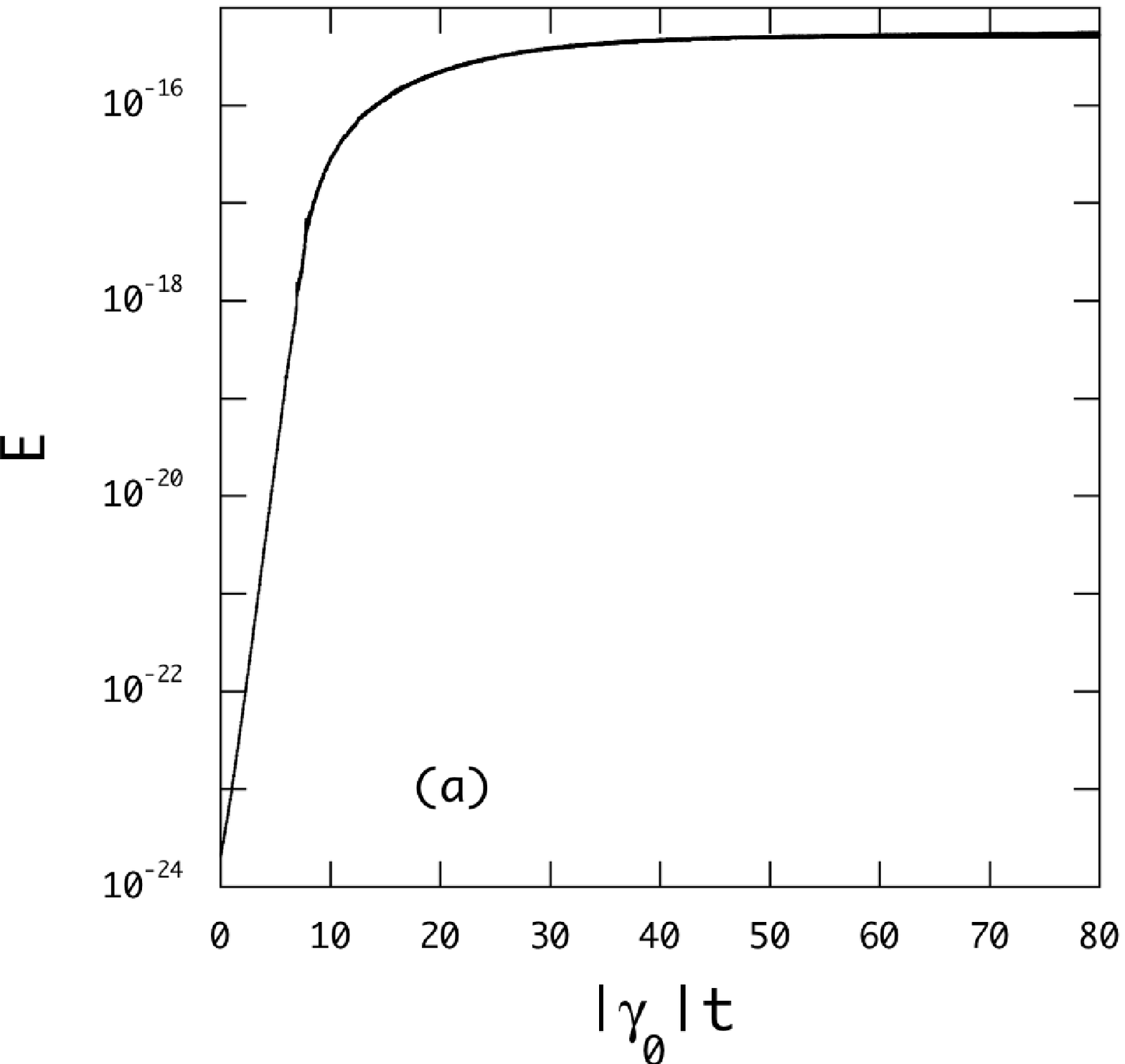}}
\resizebox{0.66\columnwidth}{!}{
\includegraphics{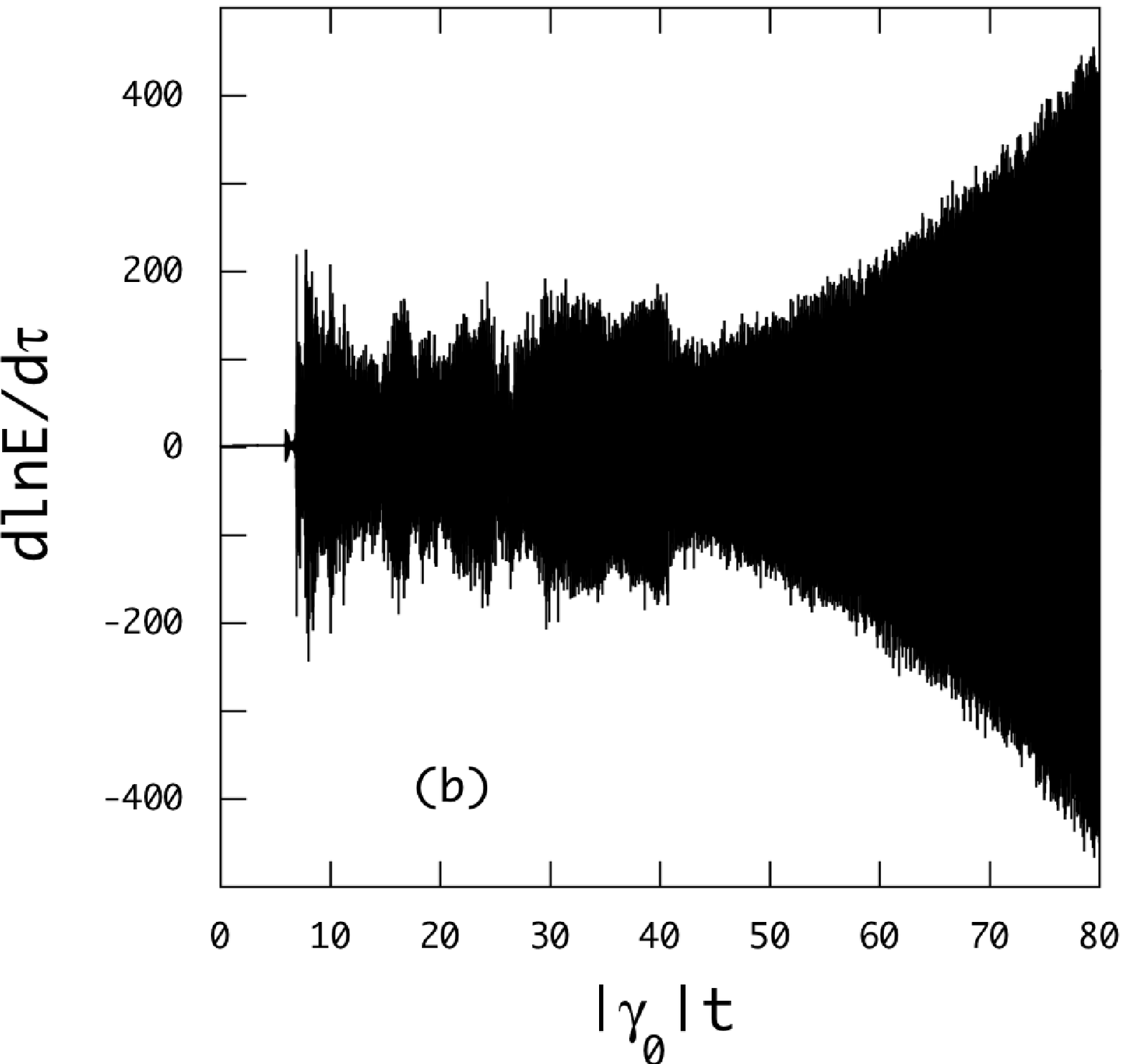}}
\resizebox{0.66\columnwidth}{!}{
\includegraphics{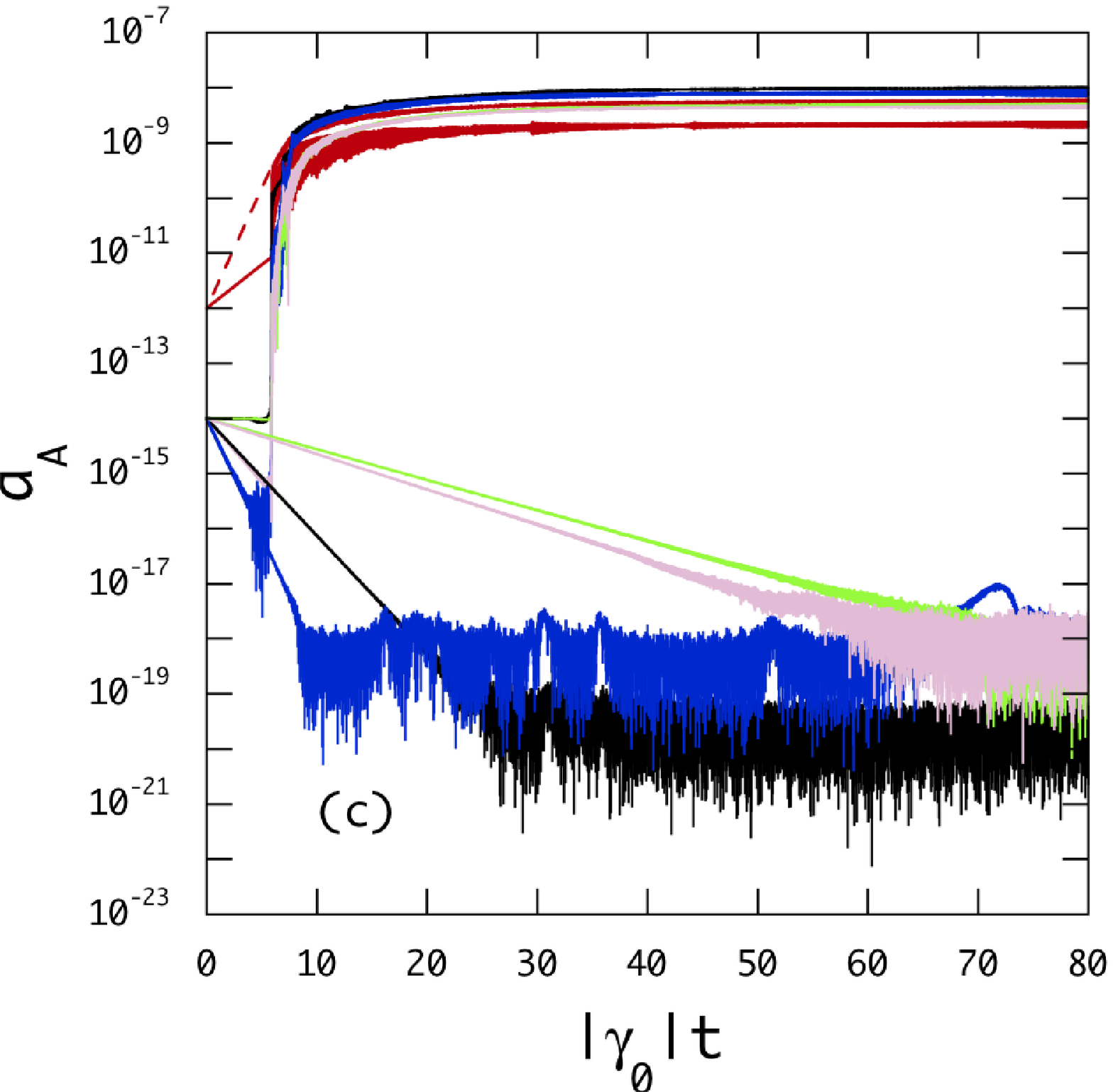}}
\caption{Same as Fig.\ref{fig:a_mm1_B1} but for the case in which the OsC modes $B_1$ and $B_2$ of $m_a=-1$
are the two parent modes where the red solid line and dashed line in panel (c) stand for the $B_1$ and $B_2$ modes,
respectively, and $|\overline\gamma_0|=1.71\times10^{-4}$.
}
\label{fig:a_mm1_B1B2}
\end{figure*}

\subsection{Non-linear Excitation and Saturation of Low Frequency Modes}

We integrate 
equations (\ref{eq:dotabc_a}) to (\ref{eq:dotabc_c}) or (\ref{eq:dotab_a}) to (\ref{eq:dotab_b})
for networks of three-mode couplings mediated by $\eta_{abc}$
to see how low frequency $g$- and $r$-modes coupled to OsC modes are excited and whether or not
the low frequency modes reach a state of finite amplitudes when $\tau\rightarrow\infty$.
Since the low radial order $g$- and $r$-modes in general have very small damping rates $\overline\gamma\ll10^{-5}$
as shown by Fig.\ref{fig:damprates}, the network inevitably contains couplings of $\Delta\gamma_{abc}<0$
between small damping rate modes and the OsC modes, 
which have $|\overline\gamma_a|\gtrsim10^{-5}$.
If the mode couplings of $\Delta\gamma_{abc}<0$ are dominant in the network of three-mode couplings,
it is likely that the evolution of mode amplitudes will fail to reach a state of finite amplitudes,
as suggested in the previous subsection.

For daughter modes, we consider $g$- and $r$-modes having the radial orders $n$ in the range
$1\lesssim n\lesssim 60$.
If the OsC modes of $m_a=-1$ are the parent modes, the low frequency daughter modes 
of $(m_b,m_c)=(-1,+2)$ and $(-2,+3)$ are computed for both even modes and odd modes.
For the OsC modes of $m_a=-2$, those of $(m_b,m_c)=(-1,+3)$ and $(+1,+1)$ are prepared.
Even (odd) modes we consider in this paper are only those whose dominant expansion coefficient of $\xi^r$, for example, tends to $S_{l=|m|}$ ($S_{l=|m|+1}$) in the limit of $\overline\Omega_s\rightarrow 0$. 
The number of daughter modes coupled to the OsC modes is $\sim 750$ for $m_a=-1$ and $\sim 700$ for $m_a=-2$.
When we integrate the amplitude equations, 
we use the initial conditions given by $a_i=10^{-12}$ for the parent modes and $b_j=c_k=10^{-14}$ for the daughter modes at $\tau=0$.

Fig. \ref{fig:a_mm1_B1} plots $E=|a_1|^2+\sum_j|b_j|^2+\sum_k|c_k|^2$, $d\ln E/d\tau$, and 
the mode amplitudes $a_A$ as a function of $\tau=|\gamma_0|t$ for $g$- and $r$-modes and the $m_a=-1$ OsC mode $B_1$
where the red line stands for the parent OsC mode and the lines of other colors are for the daughter modes of large and small amplitudes.
Note that we have used the growth rate $\overline\gamma_0=-1.71\times10^{-4}$ of the $m=-1$ OsC mode $B_2$
for normalization.
The quantity $E$ initially grows almost exponentially with $\tau$ and soon get saturated to have finite magnitudes.
Since the parent mode is initially dominant over daughter modes, we have $d\ln E/d\tau\approx -2\gamma_a/|\gamma_0|=0.726$ with $\gamma_a$ being the growth rate of
the OsC mode $B_1$ before the daughter modes are excited to 
have amplitudes comparable to the parent mode.
At the phase of daughter mode excitation, $d\ln E/d\tau$ start large amplitude fluctuations
but after the excitation is completed its fluctuation amplitude quickly decays.
As shown by the panel (c), the mode amplitudes initially grow or decay almost exponentially with time $\tau$, 
depending on whether the modes are parent or daughter modes.
When the amplitude of the parent mode
exceeds a threshold, some daughter modes are parametrically excited and their amplitudes rise abruptly
to become comparable to the parent mode while the others simply decay 
to some lower limits of amplitudes.
Among the parametrically excited daughter modes
the small damping rate modes show short period and small amplitudes fluctuations and
the large damping rate modes, on the other hand, show long period fluctuations as a function of $\tau$.
We note that the parent mode does not always obtain largest amplitudes among the low frequency modes.

In Fig.\ref{fig:faaa_B1_mm1} we plot the surface amplitudes $f_Aa_A\approx |\xi^r/R|$ of the low frequency modes
evaluated at $\tau=100$ where the red filled circle indicates
the $m_a=-1$ OsC mode $B_1$.
This figure shows that even $r^{+3}_n$-modes around the $r_{23}^{+3}$-mode 
and even $g^{-2}_n$-modes around the $g^{-2}_{58}$-mode are excited to have large amplitudes
because the coupling coefficient $\eta_{abc}$ is largest exceeding $10^8$
when the $m_a=-1$ OsC mode $B_1$ is coupled to the $r^{+3}_{23}$-mode
and the $g^{-2}_{58}$-mode.
A high peak of the amplitude $f_Aa_A$ occurs at $\overline\sigma_{c{\rm R}}\approx -0.5501$
corresponding to the $r^{+3}_{23}$-mode and another high peak at $\overline\sigma_{c{\rm R}}\approx 0.5$ to the $g^{-2}_{56}$-mode.
We note that the surface amplitudes $\xi^r/R$ of the excited modes are not necessarily very large, at most $\sim 10^{-8}$.
The surface amplitudes of the horizontal component of $\pmb{\xi}$ are by a factor $\sim 1/\overline\omega^2$
larger than the radial component and may be estimated to be of order of $\sim 10^{-6}$ for $\overline\omega\sim 0.1$ where $\overline\omega$ is the frequency in the local co-rotating frame at the surface.
The oscillations of relative amplitudes $\sim 10^{-6}$ might be difficult to detect.

In Fig. \ref{fig:a_mm1_B2} we plots $E$, $d\ln E/d\tau$, and $a_A$ for  
the $m_a=-1$ OsC mode $B_2$ as the parent mode where the growth rate of the $B_2$ mode is larger by a factor $\sim 3$ than that of the $B_1$ mode.
The sets of daughter modes we use for the $B_2$ mode are the same as those for the $B_1$ mode.
The initial behaviors of the quantities  
are quit similar to those obtained for the OsC mode $B_1$.
But, $E$ continues to increase though slowly after the excitation of daughter modes is completed and
experiences even the second rise, starting from $\tau\sim 70$.
This second rise of $E$ is accompanied by short period and large amplitude fluctuations
of $d\ln E/d\tau$.
This behavior of the quantities $E$ and $d\ln E/d\tau$ suggests that the amplitude evolutions may 
fail to reach a state of finite amplitudes.
Since the growth rate of the $B_2$ mode is by a factor $\sim3$ larger that that of the $B_1$ mode,
the effects of the three-mode couplings of $\Delta \gamma_{abc}<0$ on the amplitude evolutions
may be more significant for the former than for the latter and hence the amplitude evolutions for
the $B_2$ mode will be more unstable that for the $B_1$ mode.
Note that the even $r^{+3}_{23}$-mode has the largest amplitude after the daughter mode excitation.

We have also computed amplitude evolutions assuming that the parent mode is
the OsC mode $B_1$ of $m_a=-2$ having the growth rate $\overline\gamma_a=-1.51\times10^{-4}$, where
the daughter modes we prepare are $g$- and $r$-modes of $(m_b,m_c)=(-1,+3)$ and $(+1,+1)$.
Because the growth rate of the $B_1$ mode is large,
there are many three-mode couplings of $\Delta\gamma_{abc}<0$ and 
the amplitude evolutions fail to attain a state of finite amplitudes.

Fig. \ref{fig:a_mm1_B1B2} shows the case in which
the OsC modes $B_1$ and $B_2$ of $m_a=-1$ are the two parent modes, for which
the same mode sets as those for the parent $B_1$ mode or $B_2$ mode are used for daughter modes.
The red lines in the panel (c) stand for the parent modes $B_1$ (solid line) and $B_2$ (dashed line).
Some daughter modes are excited to obtain amplitudes comparable to the parent modes and the others
decay to lower limits of amplitudes.
Even after the daughter mode excitation is completed,
$E$ keeps increasing, though very gradually, and $d\ln E/d\tau$ starts to
fluctuate with growing amplitudes.
This suggests that the amplitude evolutions may not reach
a state of finite amplitudes.

To understand the amplitude evolutions for a given network of three-mode couplings, it may be useful
to compute the ratio $q_a$ defined by
\be
q_a= -{1\over 2}{\sum_{b,c}\left(\Delta\gamma_{abc}-\left|\Delta\gamma_{abc}\right|\right)|\eta_{abc}|\over
\sum_{b,c}\left|\Delta\gamma_{abc}\eta_{abc}\right|},
\ee
which is a measure for the relative importance of the couplings of $\Delta\gamma_{abc}<0$ in the evolutions.
For the $m_a=-1$ OsC modes we find $q_a=1.48\times10^{-2}$ for $B_1$ and $2.21\times10^{-1}$ for $B_2$ and
for the $m_a=-2$ OsC mode $B_1$ we find $q_a=4.02\times10^{-1}$.
The ratio $q_a$ is small for the network of the $m_a=-1$ OsC mode $B_1$, compared to the ratios for the other networks, and this small $q_a$ may be consistent with the amplitude evolutions that reach a state of finite amplitudes.
It will be helpful if there exists a quantity which we can use for correctly predicting the asymptotic behaviors of amplitude evolutions for a given network of mode couplings.

\section{Conclusions}

We have discussed amplitude evolutions of low frequency $g$- and $r$-modes 
excited non-linearly through three-mode couplings by the OsC modes in the core of the $2M_\odot$ ZAMS star.
Here, we have considered the lowest order non-linear couplings between three oscillation modes, one unstable and two stable modes, that is, the OsC modes for unstable parent modes and 
$g$- and $r$-modes for stable daughter modes.
We have employed the formulation given by \citet{Schenk_etal2001} to calculate the coupling coefficients between three oscillation modes where the eigenfrequencies and eigenfunctions used for the coefficients
are those of adiabatic modes and are computed using series expansions for the perturbations.
The damping rates of the $g$- and $r$-modes, however, are obtained by non-adiabatic mode calculations.
We have followed the time evolution of the amplitudes of the $g$-, $r$-, and OsC modes by integrating the 
amplitude equations to see how the $g$- and $r$-modes are excited and the OsC modes are saturated and whether or not the amplitude evolutions reach a state of finite amplitudes.
We have found that when the parent modes have small growth rates so that $\Delta\gamma_{abc}>0$ for
most of the couplings the amplitude evolutions are likely to reach a state of finite amplitudes.
On the other hand,
if the growth rates are large so that $\Delta\gamma_{abc}<0$ for many couplings with large $|\eta_{abc}|$, the evolutions may fail to tend towards states of finite amplitude.
Since the mode sets we prepared for daughter modes are quite limited,
the conclusions stated above might change, for example, if we can use much larger mode sets for daughter modes having large damping rates. 
For example, 
if we can include much higher radial order $g$- and $r$-modes with $n\gg 60$ in the daughter mode sets,
the high radial order modes with large damping rates could contribute to stabilization of the amplitude evolutions even if the parent modes have large growth rates.

Fig.\ref{fig:faaa_B1_mm1} is an example of a state of finite amplitudes obtained as a result of the amplitude evolutions of low frequency modes, which are excited by the OsC mode $B_1$ of $m_a=-1$.
This figure shows that the OsC mode can effectively excite even $r^{+3}$- and $g^{-2}$-modes around
the $r_{23}^{+3}$- and $g_{58}^{-2}$-modes, for which $|\eta_{abc}|$ takes the largest value.
Note that the other modes like $g^{+3}$-modes are not necessarily strongly excited and have much smaller amplitudes.
The mode excitation through non-linear couplings can be a possible excitation mechanism for
$r$-modes detected in many rotating stars (e.g., \citealt{Saio_etal2018}).

We have shown some examples of amplitude evolutions of low frequency modes 
non-linearly excited by the OsC modes in the core of the $2M_\odot$ ZAMS star
for $\overline\Omega_s=0.2$.
It will be important to see how parametric excitation and amplitude saturation of the low frequency modes
in rotating stars depend on the stellar mass, evolutionary stage, rotation rate, and so on.
We have found that there are cases in which the parametric excitations of low frequency modes
by OsC modes are not always sufficient to achieve saturation of the mode amplitudes.
As suggested in the earlier paragraph in this section, one possible remedy for unstable evolutions may be to include high radial order $g$- and $r$-modes
with large damping rates in the sets of daughter modes. However, these unstable amplitude evolutions
might also suggest that we have to include higher order non-linear terms in amplitude equations to obtain a state of finite amplitudes (e.g., \citealt{GoupilBuchler1994}).
This suggestion is in some sense obvious if we consider OsC modes well confined into the core.
The strongly confined OsC modes have no chance to non-linearly couple to
stable low frequency modes in the envelope for amplitude saturation.
Hence, if we assume that amplitude saturation of the OsC modes in the core
is caused by non-linear effects, we need to consider
higher order non-linear effects on the amplitude evolutions.

\section*{Acknowledgements}

The author thanks the anonymous referee for his/her comments that are of great help 
to improve the presentation of the paper.

\bigskip
\bigskip
\noindent
Data Availability: The data underlying this article will be shared on reasonable request to the corresponding author.

\bibliographystyle{mnras}
\bibliography{myref}

\end{document}